\documentclass[prd,nofootinbib,english]{revtex4}
\usepackage{amsmath}
\usepackage{amsfonts,color}
\usepackage{amsmath}
\usepackage{amsthm}
\usepackage{pdfpages}
\usepackage{amsfonts,color,xcolor}
\usepackage{amssymb,float}
\usepackage{accents}
\usepackage{hyperref}
\hypersetup{
    colorlinks=true,
    linkcolor=blue,
    filecolor=magenta,      
   citecolor=blue
}
\usepackage[utf8]{inputenc}
\allowdisplaybreaks

\newcommand{\udt}[3]{#1^{#2}_{\phantom{#2}#3}}

\newcommand{\dut}[3]{#1_{#2}^{\phantom{#2}#3}}

\newcommand{\lc}[1]{\accentset{\circ}{#1}}

\newcommand{\dd}{{\rm d}}

\newtheorem{theorem}{Theorem}
\usepackage{bookmark}

\begin{document}

\title{Black Holes in $f(T,B)$ Gravity: Exact and Perturbed Solutions}

\author{Sebastian Bahamonde}
\email{sbahamonde@ut.ee, sebastian.beltran.14@ucl.ac.uk}
\affiliation{Laboratory of Theoretical Physics, Institute of Physics, University of Tartu, W. Ostwaldi 1, 50411 Tartu, Estonia}
\affiliation{Department of Physics, Tokyo Institute of Technology
1-12-1 Ookayama, Meguro-ku, Tokyo 152-8551, Japan.}

\author{Alexey Golovnev}
\email{agolovnev@yandex.ru}
\affiliation{Centre for Theoretical Physics, British University in Egypt, 11837 El Sherouk, Cairo, Egypt}

\author{María-José Guzmán}
\email{mjguzman@ut.ee}
\affiliation{Laboratory of Theoretical Physics, Institute of Physics, University of Tartu, W. Ostwaldi 1, 50411 Tartu, Estonia} 
\affiliation{Departamento de Física y Astronomía, Facultad de Ciencias, Universidad de La Serena,
Av. Juan Cisternas 1200, 1720236 La Serena, Chile}

\author{Jackson Levi Said}
\email{jackson.said@um.edu.mt}
\affiliation{Institute of Space Sciences and Astronomy, University of Malta, Msida, Malta}
\affiliation{Department of Physics, University of Malta, Msida, Malta}

\author{Christian Pfeifer}
\email{christian.pfeifer@zarm.uni-bremen.de}
\affiliation{ZARM, University of Bremen, 28359 Bremen, Germany.}

\begin{abstract}
Spherically symmetric solutions of theories of gravity built one fundamental class of solutions to describe compact objects like black holes and stars. Moreover, they serve as starting point for the search of more realistic axially symmetric solutions which are capable to describe rotating compact objects. Theories of gravity that do not possess spherically symmetric solutions which meet all observational constraints are easily falsified. In this article, we discuss classes of exact and perturbative spherically symmetric solutions in $f(T,B)$-gravity. The perturbative solutions add to the ones which have already been found in the literature, while the exact solutions are presented here for the first time. Moreover, we present general methods and strategies, like generalized Bianchi identities, to find spherically solutions in modified teleparallel theories of gravity.
\end{abstract}

\maketitle

\section{Introduction\label{intro}}
The success of general relativity (GR) as a fundamental theory of gravity has been shown through the decades at all scales where observations are possible \cite{dodelson2003modern,Weinberg:1972kfs}. In cosmology, its amalgamation with cold dark matter (CDM) and a cosmological constant $\Lambda$ has produced the $\Lambda$CDM concordance model which meets many of the most important observational measurements from cosmology \cite{Riess:1998cb,Perlmutter:1998np} as well as observations from galactic dynamics \cite{alma991010731099705251}. On the other hand, the challenge from new observations has led to a growing disparity in a number of crucial measurements such as the value of the Hubble constant \cite{Bernal:2016gxb,DiValentino:2020zio,DiValentino:2021izs} as well as the increasing tension in the growth of large scale structure \cite{Aghanim:2018eyx,DiValentino:2020vvd} in the Universe. This has also brought to a fore internal problems in the $\Lambda$CDM model such as the cosmological constant value problem \cite{Weinberg:1988cp,Appleby:2018yci} as well as the lack of observational evidence for a particle physics explanation of CDM \cite{Baudis:2016qwx,Bertone:2004pz}.

To confront these developing problems in $\Lambda$CDM one possible recourse which is receiving an increasing attention in the literature is that of modified theories of gravity \cite{Clifton:2011jh,Capozziello:2011et,CANTATA:2021ktz}. While these theories may confront the observational tensions in $\Lambda$CDM, it is imperative that their astrophysical front also be probed for signs of phenomenology beyond GR. It is crucial to explore which of these models can offer a viable phenomenology in the context of exact and perturbative solutions, and what they may predict for the increasingly important field of black hole (BH) physics \cite{Barack:2018yly}. The precision in this field is increasingly becoming a possible discriminator between the plethora of models in modified gravity. In addition to precise observations of the final phases of the coalescence of binary BHs \cite{LIGOScientific:2016aoc}, neutron stars \cite{LIGOScientific:2017vwq}, and other scenarios \cite{LIGOScientific:2021qlt,Wang:2021iwp}, we also have confirmed observations from multimessenger astronomy \cite{LIGOScientific:2017vwq,Goldstein:2017mmi} and direct observations of the shadow of a supermassive BH \cite{EventHorizonTelescope:2019dse}. This process is set to drastically increase with the next generation of gravitational wave observatories such as LIGO India \cite{Saleem:2021iwi}, the Einstein Telescope \cite{Maggiore:2019uih} and LISA \cite{Maselli:2021men}, among others, which will drastically increase the number of event observations enlarging the possibility of observing more exotic phenomena. These observations have already had a big impact on limiting the number of viable models in certain modified theories of gravity \cite{Ezquiaga:2018btd}.

The combination of tensions in recent cosmological observations, the possibility of vast amounts of new gravitational wave measurements, and the theoretical challenges in the vacuum energy of the cosmological constant, give robust motivation for the consideration of novel approaches to constructing gravitational theories distinct to GR. One of these transformational theories of gravity is teleparallel gravity (TG) which represents those theories in which the curvature associated with the Levi-Civita connection is exchanged with the torsion linked with the teleparallel connection \cite{Bahamonde:2021gfp,Aldrovandi:2013wha,Cai:2015emx,Krssak:2018ywd}. In GR the Riemann tensor $\udt{\lc{R}}{\alpha}{\beta\mu\nu}$ (over-circles represent quantities calculated with the Levi-Civita connection) acts as a measure of curvature in a particular geometry and is wholly dependent on the Levi-Civita connection \cite{misner1973gravitation}. By exchanging this with the teleparallel connection renders a Riemann tensor that is identically zero \cite{Aldrovandi:2013wha}, $R^\alpha{}_{\beta\mu\nu} \equiv 0$. Thus, we require an analogous geometric object encoding the gravitational interaction, which corresponds to the so-called torsion tensor \cite{Bahamonde:2021gfp} computed in terms of the teleparallel connection. A mathematical identity relates the Ricci scalar $\lc{R}$ with the torsion scalar $T$ (made up of contractions of the torsion tensor) up to a boundary term $B$, guaranteeing  that a teleparallel equivalent of general relativity (TEGR) exists, which is dynamically equivalent to GR at the level of the equations of motion.

Although given this dynamical equivalence between TEGR and GR, integrating out the boundary term $B$ from the action means that the teleparallel generalization of Lovelock's theorem \cite{Lovelock:1971yv} immediately becomes much softer in that a much larger class of purely gravitational theories can produce second order field equations \cite{Gonzalez:2015sha,Bahamonde:2019shr}. TEGR is amenable to many of the possible avenues of generalization that have become well known as popular forms of generalizations of GR. As in $f(\lc{R})$ \cite{Sotiriou:2008rp,Faraoni:2008mf,Capozziello:2011et}, we can directly generalize TEGR to $f(T)$ gravity \cite{Ferraro:2006jd,Ferraro:2008ey,Bengochea:2008gz,Linder:2010py,Chen:2010va,Bahamonde:2019zea} which forms a large portion of the literature on modified TG. Due to the division of the regular Ricci scalar into the torsion scalar and boundary term, a more representative generalization of $f(\lc{R})$ gravity may be $f(T,B)$ gravity \cite{Escamilla-Rivera:2019ulu,Escamilla-Rivera:2021xql,Bahamonde:2015zma,Capozziello:2018qcp,Bahamonde:2016grb,Paliathanasis:2017flf,Farrugia:2018gyz,Bahamonde:2016cul,Wright:2016ayu,Caruana:2020szx} which limits to $f(\lc{R})$ when $f(T,B) = f(-T+B) = f(\lc{R})$. In general, $f(T,B)$ gravity has shown promise in the literature and also offers the possibility of proposing novel models of gravity that cannot appear in curvature-based forms of gravity. Another intriguing possible direction to TG is that of nonminimal coupling with scalar fields. This has been applied in various forms \cite{Bahamonde:2020vfj,Bahamonde:2019gjk,Bahamonde:2018miw,Bahamonde:2017ifa} with the most general form a teleparallel scalar field theories being formed in Refs.~\cite{Hohmann:2018rwf,Hohmann:2018vle,Hohmann:2018dqh,Hohmann:2018ijr}. Another interesting recent development is the formation of a teleparallel analogue of Horndeski gravity \cite{Bahamonde:2019shr} which is the most general theory formed by the addition of a single scalar field which turns out to produce a much larger plethora of models as compared to its curvature-based origins. This richer landscape means that a more nuanced phenomenology will emerge from this form of TG \cite{Bahamonde:2019ipm,Bahamonde:2020cfv,Bahamonde:2021dqn,Bernardo:2021bsg,Bernardo:2021izq}.

This work continues the study of finding spherically symmetric and black hole solutions and exploring their physical properties within the various scenarios available in TG \cite{Paliathanasis:2014iva,Ruggiero:2015oka,DeBenedictis:2016aze,Flathmann:2019khc,Bahamonde:2020vpb,Pfeifer:2021njm,Golovnev:2021htv,Bohmer:2019vff,Boehmer:2020hkn}. We open in Sec.~\ref{sec:intro_TG} by briefly introducing the basis on which TG is founded upon, as well as some of the many possible directions in which TEGR has been modified and the properties of the Bianchi identity in TG. In Sec.~\ref{sec:sp_sym} we consider a spherically symmetric setting in which we present the field equations for $f(T,B)$ gravity together with the possible solutions for the tetrad fields.
The specific dynamical equations for this background are then presented in Sec.~\ref{sec:pert_spherical_sym} where we also present a strategy to solving them by taking a perturbation about a Schwarzschild background. Here, we indeed arrive at specific perturbation solutions for the metric components. Interestingly, Sec.~\ref{sec:complex_tetrad} is one of the first works that explores the possibility of complex solutions in the tetrad field which continue to provide real solutions for the metric. These turn out to be somewhat simpler to solve in certain circumstances. Sec.~\ref{sec:phenomenology} then explores the phenomenological predictions from these solutions which is crucial to understanding how viable the Lagrangian models are. In Sec.~\ref{sec:time_depend} we briefly discuss the prospect of find a time dependent solution in TG, while in Sec.~\ref{sec:conclusion}, we conclude with a short discussion about our core results. In this work, we use the $(+,-,-,-)$ signature convention so that the Minkowski metric is written as $\eta_{AB} = {\rm diag}(+,-,-,-)$. Also, we use units in which $G=c=1$.

\section{\texorpdfstring{$f(T,B)$}{} teleparallel theories of gravity}\label{sec:intro_TG}
GR can be modified in a plethora of different ways \cite{Clifton:2011jh}, some of which are motivated by the Lovelock theorem but others being entirely novel in nature. TG is built on the exchange of curvature-based geometries with torsional ones \cite{Hohmann:2019nat} which takes place through replacement of the Levi-Civita and teleparallel connections. This can then develop into a myriad of modified teleparallel theories of gravity.

Curvature is expressed through the Levi-Civita connection $\lc{\Gamma}^{\sigma}_{\mu\nu}$ (over-circles are used throughout to denote quantities determined using the Levi-Civita connection) in GR, while in TG this is replaced by the teleparallel connection $\Gamma^{\sigma}_{\mu\nu}$, which is curvature-less \cite{Hayashi:1979qx,Aldrovandi:2013wha}. An immediate by-produce of this exchange is that using the teleparallel connection will render all curvature-based quantities identically zero such as the teleparallel Riemann tensor (the regular Levi-Civita connection Riemann tensor naturally does not vanish). In this context, we require a different formulation of gravitational tensors to further build gravitational theories (see reviews in Refs. \cite{Krssak:2018ywd,Cai:2015emx,Aldrovandi:2013wha}).

Another aspect of TG is that the metric $g_{\mu\nu}$ is replaced with the tetrad $\udt{h}{A}{\mu}$ (and its inverses $\dut{h}{A}{\mu}$) as the fundamental dynamical variable of the theory. This is achieved through the relations
\begin{align}\label{metric_tetrad_rel}
    g_{\mu\nu}=\udt{h}{A}{\mu}\udt{h}{C}{\nu}\eta_{AC}\,,& &\eta_{AC} = \dut{h}{A}{\mu}\dut{h}{C}{\nu}g_{\mu\nu}\,,
\end{align}
where Latin indices refer to tangent space coordinates while Greek ones refer to coordinates on the general manifold \cite{Hayashi:1979qx}. Tetrads can also be used in GR but they are generally suppressed, for example consider the description of spinors \cite{Chandrasekhar:1984siy}. Tetrads also must observe orthogonality conditions for internal consistency
\begin{align}
    \udt{h}{A}{\mu}\dut{h}{C}{\mu}=\delta^A_C\,,&  &\udt{h}{A}{\mu}\dut{h}{A}{\nu}=\delta^{\nu}_{\mu}\,.
\end{align}

The teleparallel connection is defined by \cite{Weitzenbock1923,Krssak:2018ywd}
\begin{equation}
    \Gamma^{\sigma}_{\nu\mu}:= \dut{h}{A}{\sigma}\partial_{\mu}\udt{h}{A}{\nu} + \dut{h}{A}{\sigma}\udt{\omega}{A}{D\mu}\udt{h}{D}{\nu}\,,
\end{equation}
where $\udt{\omega}{A}{D\mu}$ represents a flat spin metric-compatible connection whose components can be written as \cite{Krssak:2018ywd}
\begin{equation}
    \omega^A{}_{D\mu}=\Lambda^A{}_C\, \partial_\mu (\Lambda^{-1})^C{}_{D}\,,
\label{eq: local Lorentz transformation}
\end{equation}
where the dependence on Lorentz matrices  $\udt{\Lambda}{A}{C}$ is made explicit. This connection is responsible for incorporating the local Lorentz transformation degrees of freedom into TG, and which arises due to the explicit appearance of the Latin indices in theory. In GR, the spin connection are not flat and indeed embody elements of the non-inertial frames \cite{misner1973gravitation}. 

In parallel as the Levi-Civita connection builds up to the Riemann tensor in curvature-based gravity $\udt{\lc{R}}{\alpha}{\beta\mu\nu}$, TG is formulated using the torsion tensor \cite{Hayashi:1979qx}
\begin{equation}
    \udt{T}{A}{\mu\nu} := -2\Gamma^{A}_{[\mu\nu]} = \partial_{\mu}h^A{}_{\nu}-\partial_{\nu}h^A{}_{\mu} + \omega^A{}_{B\mu} h^B{}_{\nu}-\omega^A{}_{B\nu} h^B{}_{\mu}\,,
\end{equation}
where square brackets denote an antisymmetric operator. The torsion tensor is sometimes interpreted as the gauge field strength of gravity in TG \cite{Aldrovandi:2013wha}. The torsion tensor transforms covariantly under both diffeomorphisms and local Lorentz transformations. Taking suitable contractions leads to the torsion scalar given by \cite{Krssak:2018ywd,Cai:2015emx,Aldrovandi:2013wha}
\begin{equation}
    T=\frac{1}{2}S^{\alpha\mu\nu}T_{\alpha \mu \nu}=\frac{1}{4}T^{\mu\nu\rho}T_{\mu\nu\rho} + \frac{1}{2}T^{\mu\nu\rho}T_{\rho\nu\mu} - T_{\rho}T^{\rho}\,,\label{defT}
\end{equation}
where $T_{\rho} := \udt{T}{\alpha}{\rho\alpha}$, and where $\dut{S}{\rho}{\mu\nu}$ is the superpotential tensor described by
\begin{eqnarray}
S_{\rho}{}^{\mu\nu} &=&K^{\mu\nu}{}_{\rho}-\delta_{\rho}^{\mu}T_{\sigma}{}^{\sigma\nu}+\delta_{\rho}^{\nu}T_{\sigma}{}^{\sigma\mu}=-S_{\rho}{}^{\nu\mu}\,,
\end{eqnarray}
and where the contortion tensor $\udt{K}{\rho}{\mu\nu}$ is given by 
\begin{eqnarray}
K^{\rho}{}_{\mu\nu} & =&\Gamma^{\rho}{}_{\mu\nu}-\lc{\Gamma}^{\rho}{}_{\mu\nu}=\frac{1}{2}\left(T_{\mu}{}^{\rho}{}_{\nu}+T_{\nu}{}^{\rho}{}_{\mu}-T^{\rho}{}_{\mu\nu}\right)\,,
\end{eqnarray}
which plays an important role in relating TG with curvature-based theory analogues. 

Using the contortion tensor and simplifying, the curvature-based Ricci scalar $\lc{R}$ can be shown to be expressed through \cite{Bahamonde:2015zma,Farrugia:2016qqe}
\begin{equation}\label{LC_TG_conn}
    R =\lc{R} + T - \frac{2}{h}\partial_{\mu}\left(h T^{\sigma}{}_{\sigma\mu}\right) = 0\,,
\end{equation}
where $R$ identically vanishes since it is torsion-less, and where the last term is a boundary term. Thus, we can write
\begin{equation}\label{TEGR_L}
R=\lc{R} +T-\frac{2}{h}\partial_{\mu}\left(hT^{\sigma}{}_{\sigma\mu}\right)=0 \quad \Rightarrow \quad \lc{R} = -T + \frac{2}{h}\partial_{\mu}\left(hT^{\sigma}{}_{\sigma\mu}\right) := -T + B\,,
\end{equation}
where it becomes clear that by replacing the Ricci scalar in the Einstein-Hilbert action with the linear torsion will produce a dynamically identical theory. This is the teleparallel equivalent of general relativity (TEGR).

\subsection{Action and field equations}
\label{sec:action}

Nonlinear extensions to the TEGR Lagrangian $\mathcal{L}=hT$ are motivated by following the recipe for other extended theories of gravity, such as $f(\lc{R})$ gravity \cite{DeFelice:2010aj,Capozziello:2011et}. Although the simplest extension is the so-called $f(T)$ gravity, which has been widely studied in the literature, in this work we extend a bit further and define $f(T,B)$ gravity, given by the following action
\begin{align}
\mathcal{S}_{ f(T,B)} = \int d^4x\, h\, \left[ \frac{1}{2\kappa^2}f(T,B) + \mathcal{L}_{\text{m}}
\right] \,,\label{action}
\end{align}
where $\kappa^2=8\pi G$, $\mathcal{L}_{\text{m}}$ is the matter Lagrangian, and $h=\det\left(\udt{h}{a}{\mu}\right)=\sqrt{-g}$ is the tetrad determinant. The TEGR limit is surely included in this class of theories and corresponds to $f(T,B) = -T$. 

The ensuing field equations for this theory turn out to be \cite{Bahamonde:2015zma}
\begin{align}
  \dut{W}{\nu}{\lambda} := \delta_{\nu}^{\lambda}\lc{\Box} f_{B} &- \lc{\nabla}^{\lambda}\lc{\nabla}_{\nu}f_{B}+
  \frac{1}{2}B f_{B}\delta_{\nu}^{\lambda} + 
  \Big[(\partial_{\mu}f_{B})+(\partial_{\mu}f_{T})\Big]S_{\nu}{}^{\mu\lambda}\nonumber\\
  &+ h^{-1}h^{A}{}_{\nu}\partial_{\mu}(h S_{A}{}^{\mu\lambda})f_{T} - 
  f_{T}T^{\sigma}{}_{\mu \nu}S_{\sigma}{}^{\lambda\mu} - \frac{1}{2}
  f \delta_{\nu}^{\lambda}
  =\kappa^2 \dut{\mathcal{T}}{\nu}{\lambda}\,,\label{fieldeq}
\end{align}
where subscripts denote derivatives, and where $\dut{\mathcal{T}}{\nu}{\lambda}$ represents the energy-momentum tensor. The separate tetrad and spin connection variations produce the field equations
\begin{equation}
    W_{(\mu\nu)} = \kappa^2 \Theta_{\mu\nu}\,, \quad \text{and} \quad W_{[\mu\nu]} = 0\,,\label{FE_symm}
\end{equation}
which are respectively associated with the tetrad and spin connection degrees of freedom. 

Given any metric tensor, a unique tetrad-spin connection pair exists in which the local frame is compatible with a spin connection whose components all vanish, which is called the Weitzenb\"{o}ck gauge \cite{Krssak:2018ywd}. In this setting, $W_{[\mu\nu]}$ vanishes identically due to the choice of the tetrad components (while still satisfying the metric relation~(\ref{metric_tetrad_rel})). For the action in Eq.~(\ref{action}), the antisymmetric equations can be succinctly written as
\begin{equation}
  W_{[\mu\nu]} \equiv \Big[(\partial_{\rho}f_{B})+(\partial_{\rho}f_{T})\Big]S_{[\mu}{}^{\rho}{}_{\nu]}\propto T^{\rho}{}_{[\mu\nu}\partial_{\rho]}(f_T+f_B)\,,\label{antitotal}
\end{equation}
which will be used to determine tetrad solutions of this kind in the work that follows.

\subsection{Bianchi identities}\label{ssec:BI}

Since our aim throughout this work is to study Eqs. \eqref{fieldeq} in the context of spherical symmetry, it will be helpful to first discuss Bianchi identities in the context of modified teleparallel theories. An important point about spherically symmetric solutions is that the three equations of motion obtained are not fully independent of each other. In the case of GR the relation among them corresponds to the Bianchi identity $\lc{R}^\alpha{}_{\beta[\lambda\mu;\nu]}=0$, which leads to covariant conservation of the Einstein tensor. It has been shown \cite{Golovnev:2020las} that this statement can be generalised to $f(T)$ theories. Namely, if the antisymmetric part of equations is satisfied, then the covariant divergence of equations of motion vanishes identically.

Since the derivation of these generalised Bianchi identities was based simply on the diffeomorphism invariance of the action, it can be further generalised to other models. The formulation would be equivalent in one extend the theory to also depend on a scalar field and a kinetic term. 

Let us look, for simplicity, at the equations in the Weitzenb{\" ock} gauge $\omega^A{}_{\mu\nu}=0$. The diffeomorphism invariance of the action is then invariance under
\begin{equation}
h^A_{\hphantom{A}\mu}\longrightarrow h^A_{\hphantom{A}\mu}-h^A_{\hphantom{A}\nu}\partial_{\mu}\zeta^{\nu}-\zeta^{\nu}\partial_{\nu}h^A_{\hphantom{A}\mu}\,,
\end{equation}
which for the equations of motion ${\mathfrak T}^{\mu\nu}=0$ with
\begin{equation}
\kappa^2\frac{\delta {\mathcal S}}{\delta h^A_{\hphantom{A}\mu}}\equiv h {\mathfrak T}^{\mu\nu} h^B_{\hphantom{A}\nu}\eta_{AB}\,,
\end{equation}
implies the following simple relation:
\begin{equation}
\lc{\nabla}_{\mu}{\mathfrak T}^{\mu\nu}+K^{\alpha\beta\nu}{\mathfrak T}_{\alpha\beta}=0\,.
\end{equation}
Due to the antisymmetry of the contortion, it means that once the antisymmetric part of equations is satisfied then the Bianchi identities are satisfied too. 

Let us now see how it works directly at the level of equations of motion for $f(T,B)$ gravity. For the sake of simplicity, we don't go for more elaborate models since all the main features will be clearly seen in this example. It is convenient to bring the equations to a more (diffeo-)covariant form which in the case of $f(T,B)$ in vacuum reads
\begin{equation}
f_{T}\lc{G}_{\mu\nu}+\frac12 \left(f-f_{T}T-f_B B\right)g_{\mu\nu}-\left(g_{\mu\nu}\lc{\square}- \lc{\nabla}_{\mu}\lc{\nabla}_{\nu}\right)f_B+S_{\mu\nu\alpha}\partial^{\alpha}(f_T + f_B)=0\,.
\end{equation}
As also in the case of $f(T)$, only the last term is the one which contains the antisymmetric part. 

Taking the divergence of the left hand side goes similar to the computation that has been done for $f(T)$ gravity in the paper \cite{Golovnev:2020las}, though with more terms:
\begin{multline}
\label{divergence}
f_{TT}\left(\lc{G}^{\mu\nu}\partial_{\nu} T+(\partial_{\alpha} T)\lc{\nabla}_{\nu}S^{\mu\nu\alpha}-\frac12 g^{\mu\nu} T\partial_{\nu} T \right)+f_{TB}\left(\lc{G}^{\mu\nu}\partial_{\nu}B+(\partial_{\alpha}( B+ T))\lc{\nabla}_{\nu}S^{\mu\nu\alpha}-\frac12 g^{\mu\nu}( T \partial_{\nu}B+B\partial_{\nu} T)\right)\\
+f_{BB}\left((\partial_{\alpha}B)\lc{\nabla}_{\nu} S^{\mu\nu\alpha}-\frac12 B\partial_{\nu}B\right)-\left(\lc{\nabla}_{\mu}\lc{\nabla}_{\nu}-\lc{\nabla}_{\nu}\lc{\nabla}_{\mu}\right)\lc{\nabla}^{\nu}f_B\,.
\end{multline}

The first bracket is what also appears in $f(T)$. It can be seen to vanish if the antisymmetric part of $f(T)$ equations is satisfied. Indeed, recall that
$$\lc{G}^{\mu\nu}=K^{\hphantom{\alpha\rho}\mu}_{\alpha\rho}S^{\alpha\rho\nu}-\lc{\nabla}_{\alpha}S^{\mu\alpha\nu}+\frac12  T g^{\mu\nu}\,,$$
and then this first bracket above goes away if the antisymmetric part of $S^{\alpha\rho\nu}\partial_{\nu}  T$ is zero (antisymmetry of the contortion tensor $K^{\hphantom{\alpha\rho}\mu}_{\alpha\rho}$ again).

Now we have more terms, and also the new antisymmetric part of equations is $$S^{[\alpha\rho]\nu}\partial_{\nu}(f_T+f_B)=S^{[\alpha\rho]\nu}(f_{TT}\partial_{\nu} T+f_{TB}\partial_{\nu}(B+ T)+f_{BB}\partial_{\nu} B)=0\,.$$ This is what is equal to zero under the assumption of satisfaction of antisymmetric equations. Therefore the story is similar to $f(T)$ but requires a bit more accuracy.

The new step we need to do is to transform the commutator of derivatives acting on the gradient of $f_B$. It gives  $$-\left(\lc{\nabla}_{\mu}\lc{\nabla}_{\nu}-\lc{\nabla}_{\nu}\lc{\nabla}_{\mu}\right)\lc{\nabla}^{\nu}f_B=\lc{R}^{\mu\nu}(f_{TB}\partial_{\nu} T+f_{BB}\partial_{\nu}B)=\left(\lc{G}^{\mu\nu}+\frac12 g^{\mu\nu}\lc{R}\right)(f_{TB}\partial_{\nu} T+f_{BB}\partial_{\nu}B)\,.$$
Together with the identity $\lc{R}=- T + B$, see \eqref{TEGR_L}, it brings the full divergence (\ref{divergence}) of the equations of motion to the form
$$\left(\lc{G}^{\mu\alpha}+\lc{\nabla}_{\nu}S^{\mu\nu\alpha}-\frac12 g^{\mu\alpha} T \right)(f_{TT}\partial_{\alpha} T+f_{TB}\partial_{\alpha}(B+ T)+f_{BB}\partial_{\alpha} B)\,.$$
Substituting again the expression for the Einstein tensor from above, we see that its first term (with the contortion tensor) vanishes if the antisymmetric part of equations is satisfied, while the next two terms cancel everything else, therefore confirming the generalised Bianchi identity for $f(T,B)$ gravity.

\section{Anti-symmetric field equations of \texorpdfstring{$f(T,B)$}{} gravity assuming spherical symmetry\label{sec:sp_sym}}
Our aim is to solve the field equations \eqref{fieldeq} for spherically symmetric systems. Therefore we first recall the implementation for spherical symmetry in $f(T,B)$ gravity. Afterwards we solve the anti-symmetric equations.

\subsection{Spherical symmetry}
\label{sec:spherical_symm}
In teleparallel gravity spacetime symmetries, or more technically, the invariance of the geometry on under a certain class of diffeomorphisms, can be characterized infinitesimally by symmetry generating vector fields $Z_{\zeta}$ and the demand that the geometry defining fields, the tetrad and the spin connection, satisfy the following teleparallel Killing equations, see \cite{Hohmann:2019nat} for all details,
\begin{align}
\mathcal{L}_{Z_\zeta}h^{A}\,_{\mu}=-\,\lambda^{A}_{\zeta}{}_{B}h^{B}\,_{\mu} \,,\quad 
\mathcal{L}_{Z_\zeta}\omega^{A}\,_{B\mu}=\partial_{\mu}\lambda^{A}_{\zeta}{}_{B}+\omega^{A}\,_{C\mu}\lambda^{C}_{\zeta}{}_{B}-\omega^{C}\,_{B\mu}\lambda^{A}_{\zeta}{}_{C}\,.\label{tpsy}
\end{align}
The $\lambda^{A}_{\zeta}{}_{B}$ are elements of the Lie algebra $\mathfrak{so}(1,3)$ (the Lorentz algebra), which are obtained by mapping the symmetry algebra defined by the vector fields $Z_{\zeta}$ into the Lorentz algebra.

Spherical spacetime symmetry is defined by the generating vector fields 
\begin{align}
    Z_1 &= \sin \varphi \partial_\vartheta + \frac{\cos \varphi}{\tan \vartheta}\partial_\varphi\\
    Z_2 &= -\cos \varphi \partial_\vartheta + \frac{\sin \varphi}{\tan \vartheta}\partial_\varphi\\
    Z_3 &= \partial_\varphi \,,
\end{align}
which are a representation of the Lie algebra $\mathfrak{so}(3)$ of spatial rotations in three dimensions. To map the $\mathcal{so}(3)$ homemorphic into the Lorentz algebra  $\mathfrak{so}(1,3)$ we use
\begin{align}
    \lambda(Z_1) = 
    \left(
    \begin{array}{cccc}
    0 & 0 & 0 & 0\\
    0 & 0 & 0 & 0\\
    0 & 0 & 0 & -1\\
    0 & 0 & 1 & 0\\
    \end{array}
    \right),\quad
    \lambda(Z_2) = 
    \left(
    \begin{array}{cccc}
    0 & 0 & 0 & 0\\
    0 & 0 & 0 & 1\\
    0 & 0 & 0 & 0\\
    0 & -1 & 0 & 0\\
    \end{array}
    \right),\quad
    \lambda(Z_3) = 
    \left(
    \begin{array}{cccc}
    0 & 0 & 0 & 0\\
    0 & 0 & -1 & 0\\
    0 & 1 & 0 & 0\\
    0 & 0 & 0 & 0\\
    \end{array}
    \right)\,.
\end{align}

Solving the teleparallel Killing equations \eqref{tpsy} for vanishing spin connection, then yields the most general spherically symmetric tetrad in Weitzenb\"ock gauge, which can be expressed as, see again~\cite{Hohmann:2019nat},
\begin{equation}
h^A{}_{\nu}=\left(
\begin{array}{cccc}
C_1 & C_2 & 0 & 0 \\
C_3 \sin\vartheta \cos\varphi & C_4 \sin\vartheta \cos\varphi & C_5 \cos\vartheta \cos \varphi - C_6 \sin\varphi  & -\sin\vartheta (C_5 \sin\varphi + C_6 \cos\vartheta \cos\varphi)  \\
C_3 \sin\vartheta \sin\varphi & C_4 \sin\vartheta \sin\varphi & C_5 \cos\vartheta \sin \varphi + C_6 \cos\varphi  & \sin\vartheta (C_5 \cos\varphi - C_6 \cos\vartheta \sin\varphi) \\
C_3 \cos\vartheta & C_4 \cos\vartheta & - C_5 \sin\vartheta & C_6 \sin^2\vartheta\\
\end{array}
\right)\label{sphtetrad}\,,
\end{equation}
where the six free functions $C_I=C_I(t,r)$ ($I=1,..6$) can depend on  time and the radial coordinate. The metric becomes
\begin{eqnarray}\label{metric}
\dd s^2&=&\left(C_1^2-C_3^2\right)\dd t^2 -2 ( C_3 C_4-C_1 C_2)\,\dd t\,\dd r - \left(C_4^2-C_2^2\right)\dd r^2- \left(C_5^2+C_6^2\right)\dd \Omega^2\,.
\end{eqnarray}
We can still use the freedom of redefining the $t$ and $r$ coordinate in such a way that $C_5^2+C_6^2 = r^2$ and $C_3 C_4-C_1 C_2=0$.

So far, the tetrad \eqref{sphtetrad} was identified on a purely geometric basis by symmetry considerations. In the next section we will find the time-independent spherically symmetric Weitzenb\"ock tetrads which solve the anti-symmetric field equations of $f(T,B)$ gravity.

\subsection{Solving the anti-symmetric field equations}
\label{sec:good_ted_spin}
For the tetrad~(\ref{sphtetrad}) (static case) there are two non-vanishing antisymmetric field equations~(\ref{antitotal}) given by
\begin{eqnarray}
E_{[tr]}&\propto& C_3 C_5 (f'_T+f'_B)=0\,,\label{anti1}\\
E_{[\vartheta\varphi]}&\propto& C_1 C_6(f_T'+f_B')=0\,.\label{anti2}
\end{eqnarray}
Here, primes denote differentiation with respect to $r$. We will further assume that $f_T'+f_B'\neq0$, otherwise
the theory would become $f(T,B)=\tilde{f}(-T+B)=\tilde{f}(\lc{R})$, which is just the standard $f(\lc{R})$ gravity.

There are two different branches of solutions: a) $C_3=0$; b) $C_3\neq0$. For the first branch one can solve both antisymmetric field equations by taking
\begin{eqnarray}
C_3(r)=C_6(r)=0\,.
\end{eqnarray}
Since the metric still has some extra degrees of freedom, without loosing generality, one can choose $C_2=0$ to eliminate the cross term and also set 
\begin{eqnarray}
C_5(r)=\xi r\,,\quad \xi=\pm 1\,,
\end{eqnarray}
to obtain a $r^2\dd\Omega^2$ in the line-element. Then, the first branch that solves the antisymmetric field equations has the following tetrad (in the Weitzenb\"ock gauge) and metric
\begin{eqnarray}\label{tetrad1}
h_{(1)}^{A}{}_\mu&=&\left(
\begin{array}{cccc}
 \mathcal{A}(r) & 0 & 0 & 0 \\
 0 & \mathcal{B}(r) \sin\vartheta \cos\varphi & \xi  r \cos\vartheta \cos\varphi & -r\xi \sin\vartheta \sin\varphi \\
 0 & \mathcal{B}(r) \sin\vartheta \sin\varphi & \xi  r \cos\vartheta \sin\varphi & \xi  r \sin\vartheta \cos\varphi \\
 0 & \mathcal{B}(r) \cos\vartheta & -r\xi \sin\vartheta & 0 \\
\end{array}
\right)\,,\quad \xi=\pm 1\,,\\
\dd s^2&=&\mathcal{A}(r)^2\dd t^2-\mathcal{B}(r)^2\dd r^2-r^2\dd \Omega^2\,,
\end{eqnarray}
where we have re-defined the functions $C_1=\mathcal{A}(r)$ and $C_4=\mathcal{B}(r)$.
If one chooses $\xi=1$, we recover the tetrad used in~\cite{Bahamonde:2020vpb,Bahamonde:2020bbc,Bahamonde:2019jkf,Bahamonde:2019zea} whereas if $\xi=-1$ we recover the tetrad assumed in~\cite{Ruggiero:2015oka,Finch:2018gkh,Farrugia:2016xcw,Iorio:2015rla}
For this branch, the torsion scalar and the boundary term  become
\begin{eqnarray}
T&=&\frac{2 (\xi-\mathcal{B}) }{r^2 \mathcal{A} \mathcal{B}^2}\left(2 r \xi \mathcal{A}'+\mathcal{A} (\xi-\mathcal{B})\right)\,,\\
B&=&\frac{2 \mathcal{A}''}{\mathcal{A} \mathcal{B}^2}+\frac{2\mathcal{A}'}{\mathcal{A} \mathcal{B}^2} \left(\frac{2(2\xi-\mathcal{B})}{r \xi}-\frac{ \mathcal{B}'}{\mathcal{B}}\right)-\frac{4 \mathcal{B}'}{r \mathcal{B}^3}+\frac{4 (\xi-\mathcal{B})}{r^2 \xi \mathcal{B}^2}\,.
\end{eqnarray}
One notices that $\mathcal{A}$ and $\mathcal{B}$ can be positive of real without changing the form of the metric. This means that for Minkowski, one can either set $\mathcal{A}=\mathcal{B}=1$ or $\mathcal{A}=\mathcal{B}=-1$. Then, for the Minkowski metric, the torsion scalar and boundary term become
\begin{eqnarray}
T&=&\frac{2}{r^2}(\xi-1)^2\,,\quad B=\frac{4}{\xi r^2}(\xi-1)\,, \quad \mathcal{A}=\mathcal{B}=1\,,\label{eq:xi_pos}\\
T&=&\frac{2  }{r^2}(\xi+1)^2\,,\quad B=\frac{4 }{r^2 \xi }(\xi+1)\,,\quad \mathcal{A}=\mathcal{B}=-1\,,\label{eq:xi_neg}
\end{eqnarray}
so that, for the case $\xi=1$ the torsion scalar and the boundary term vanishes for the Minkowski metric only by choosing  $\mathcal{A}=\mathcal{B}=1$ and for $\xi=-1$, we require the limit $\mathcal{A}=\mathcal{B}=-1$.

For the second branch $C_3\neq0$, one can solve all the antisymmetric field equations~(\ref{anti1})-(\ref{anti2}) by taking
\begin{eqnarray}
C_1(r)=C_5(r)=0\,.
\end{eqnarray}
One needs to be careful with this choice since the signature of the metric changes its sign, unless the tetrad functions are complex. Then, we can redefine
\begin{eqnarray}
C_2(r)=i \,\mathcal{B}(r)\,,\quad C_3(r)=i \,\mathcal{A}(r)\,,
\end{eqnarray}
in order to ensure that the metric does not change its signature. As we did in the previous branch, we can still have the metric freedom to choose 
\begin{eqnarray}
C_4=0\,,\quad C_6=\chi r\,,\quad \chi=\pm 1\,.
\end{eqnarray}
Thus, the second branch which solves the antisymmetric field equations have the following tetrad and metric:
\begin{eqnarray}\label{tetrad2}
h_{(2)}^{A}{}_\mu&=&\left(
\begin{array}{cccc}
 0 & i \mathcal{B}(r) & 0 & 0 \\
 i \mathcal{A}(r) \sin\vartheta \cos\varphi & 0 &- \chi r \sin\varphi & -r\chi \sin\vartheta \cos\vartheta \cos\varphi \\
 i \mathcal{A}(r) \sin\vartheta \sin\varphi & 0 & \chi  r \cos\varphi & -r\chi \sin\vartheta \cos\vartheta \sin\varphi \\
 i \mathcal{A}(r) \cos\vartheta & 0 & 0 & \chi  r \sin^2\vartheta \\
\end{array}
\right)\,,\quad \chi=\pm 1\,,\\
\dd s^2&=&\mathcal{A}(r)^2\dd t^2-\mathcal{B}(r)^2\dd r^2-r^2\dd \Omega^2\,.
\end{eqnarray}
The above tetrad is complex and to the best of our knowledge, it has not been derived or used before in the literature. Even though the tetrad is complex, both the torsion scalar and the boundary term are real:
\begin{eqnarray}
T&=&\frac{2}{r^2 \mathcal{B}^2 \mathcal{A}} \left(\left(\mathcal{B}^2+1\right) \mathcal{A}+2 r \mathcal{A}'\right)\,,\label{eq:torsionscalarcomplex}\\
B&=&\frac{\left(8 \mathcal{B}-2 r \mathcal{B}'\right) \mathcal{A}'}{r \mathcal{B}^3 \mathcal{A}}+\frac{4 \left(\mathcal{B}-r \mathcal{B}'\right)}{r^2 \mathcal{B}^3}+\frac{2 \mathcal{A}''}{\mathcal{B}^2 \mathcal{A}}\,.
\end{eqnarray}
Moreover, these scalars do not depend on the sign of $\chi$ as it occurs in the first branch. Then, the field equations will be identical for $\chi=\pm 1$. Without loosing generality one can  then choose $\chi=1$. For the Minkowski case,  these scalars are always non-vanishing ($T=B=4/r^2$).

The tetrads~\eqref{tetrad1} and \eqref{tetrad2} were found in the Weitzenb\"ock gauge ($w^A{}_{B\mu}=0$). Since our theory is covariant under local Lorentz transformations (by taking simultaneous transformations on the tetrad and spin connection), one can also perform a local Lorentz transformation and rewrite them as diagonal tetrads with non-vanishing spin connections. For instance, by performing the followings local Lorentz transformation $\Lambda_{(1)}^A{}_B$ and $\Lambda_{(2)}^A{}_B$ to the tetrads $h_{(1)}^A{}_\mu$ and $h_{(2)}^A{}_\mu$ 
\begin{eqnarray}
\Lambda_{(1)}^A{}_B&=&\left(
\begin{array}{cccc}
 1 & 0 & 0 & 0 \\
 0 & \sin \vartheta \cos \varphi & \sin \vartheta \sin \varphi & \cos \vartheta \\
 0 & \cos \vartheta \cos \varphi & \cos \vartheta \sin \varphi & -\sin \vartheta \\
 0 & -\sin \varphi & \cos \varphi & 0 \\
\end{array}
\right)\,,\\
\Lambda_{(2)}^A{}_B&=&\left(
\begin{array}{cccc}
 0 & -i \sin \vartheta  \cos \varphi & -i \sin \vartheta \sin \varphi  & -i \cos \vartheta \\
 -i & 0 & 0 & 0 \\
 0 & -\sin \varphi & \cos \varphi  & 0 \\
 0 & -\cos \vartheta \cos \varphi  & -\cos \vartheta \sin \varphi  & \sin \vartheta \\
\end{array}
\right)\,,
\end{eqnarray}
we end up with diagonal tetrads
\begin{eqnarray}\label{tetrad3}
h_{(1)}'^{A}{}_\mu=\textrm{diag}(\mathcal{A}(r),\mathcal{B}(r),\xi r,\xi r\sin\vartheta)\,,\quad h_{(2)}'^{A}{}_\mu=\textrm{diag}(\mathcal{A}(r),\mathcal{B}(r),r, r\sin\vartheta)
\end{eqnarray}
and with non-zero spin connection $w_{(1)}'^{A}{}{}_{B\mu}$ and $w_{(2)}'^{A}{}_{B\mu}$ having the following non-zero components
\begin{eqnarray}
w_{(1)}'^{1}{}_{2\vartheta}=-w_{(1)}'^{2}{}_{1\vartheta}=-1\,,\quad w_{(1)}'^{1}{}_{3\varphi}=-w_{(1)}'^{3}{}_{1\varphi}=-\sin\vartheta\,,\quad w_{(1)}'^{2}{}_{3\varphi}=-w_{(1)}'^{3}{}_{2\varphi}=-\cos\vartheta\,, \label{spin2}
\end{eqnarray}
\begin{eqnarray}
w_{(2)}'^{0}{}_{2\varphi}=w_{(2)}'^{2}{}_{0\varphi}=i \sin\vartheta\,,\quad w_{(2)}'^{0}{}_{3\vartheta}=w_{(2)}'^{3}{}_{0\vartheta}=-i\,,\quad w_{(2)}'^{2}{}_{3\varphi}=-w_{(2)}'^{3}{}_{2\varphi}=-\cos\vartheta\,.\label{spin3}
\end{eqnarray}
It is then equivalent to take the tetrad spin connection pairs $(h_{(1)}^A{}_\mu,0)$ given by \eqref{tetrad1} or $(h_{(1)}'^{A}{}_\mu,w_{(1)}'^{A}{}_{B\mu})$ given by~\eqref{tetrad3}-\eqref{spin2}. Conversely, the dynamics of the field equations will be equivalent if one takes the pair $(h_{(2)}^A{}_\mu,0)$ (see Eq.~\eqref{tetrad2}) or $(h_{(2)}'^{A}{}_\mu,w_{(2)}'^{A}{}_{B\mu})$ (see Eqs.~\eqref{tetrad3} and \eqref{spin3}). One can notice that for the complex tetrad~\eqref{tetrad2}, it is possible to eliminate its imaginary part by performing a local Lorentz transformation, but then the imaginary terms will be induced in the spin connection~\eqref{spin3}.

One can then conclude that there are three different tetrads in the Weitzenb\"ock gauge for $f(T,B)$ gravity, Eqs.~(\ref{tetrad1}) (with $\xi=1$ or $\xi=-1$) and (\ref{tetrad2}), which satisfy the antisymmetric field equations in the time-independent spherically symmetric case. These three tetrads respect spherical symmetry as~(\ref{tpsy}). Let us remark here that if one generalizes the theory to depend on a scalar field a kinetic term like
$f(T,B,\phi,X)$, the same tetrads obtained above would solve the antisymmetric field equations. Thus, adding a scalar field does not change the conclusions made here.Hereafter, we will study the symmetric field equations for each tetrad and find different spherically symmetric solutions for $f(T,B)$ gravity.

\section{Symmetric field equations and perturbed solutions for the real tetrad}\label{sec:pert_spherical_sym}
In this section we will concentrate on the real tetrad~(\ref{tetrad1}) and analyse the remaining field equations (symmetric ones) for $f(T,B)$ gravity. If we consider an anisotropic perfect fluid, for the tetrad~(\ref{tetrad1}), the field equations~(\ref{fieldeq}) become
\begin{eqnarray}
\kappa^2\rho&=&-\frac{1}{2}f-\frac{2 f_T \left(r \mathcal{B} \mathcal{A}' (\xi  \mathcal{B}-1)+\mathcal{A} \left(r \mathcal{B}'+\xi  \mathcal{B}^2-\mathcal{B}\right)\right)}{r^2 \mathcal{A} \mathcal{B}^3}+\frac{(2-2 \xi  \mathcal{B}) f_T'}{r \mathcal{B}^2}\nonumber\\
&&+\frac{f_B \left(r \left(-r \mathcal{A}' \mathcal{B}'-2 \xi  \mathcal{B}^2 \mathcal{A}'+\mathcal{B} \left(r \mathcal{A}''+4 \mathcal{A}'\right)\right)-2 \mathcal{A} \left(r \mathcal{B}'+\xi  \mathcal{B}^2-\mathcal{B}\right)\right)}{r^2 \mathcal{A} \mathcal{B}^3}+\frac{f_B' \left(\mathcal{B}'-\frac{2 \xi  \mathcal{B}^2}{r}\right)}{\mathcal{B}^3}-\frac{f_B''}{\mathcal{B}^2}\,,\label{fieldeqsym1}\\
\kappa^2 p_r&=&\frac{1}{2}f+\frac{\left(r \mathcal{A}'+2 \mathcal{A}\right) f_B'}{r \mathcal{A} \mathcal{B}^2}+\frac{f_T \left(2 r \mathcal{A}' (\mathcal{B}-2 \xi )-2 \mathcal{A} \left(\xi-\mathcal{B}\right)\right)}{\xi  r^2 \mathcal{A} \mathcal{B}^2}\nonumber\\
&&+\frac{f_B \left(r \left(\xi  r \mathcal{A}' \mathcal{B}'+2 \mathcal{B}^2 \mathcal{A}'-\xi  \mathcal{B} \left(r \mathcal{A}''+4 \mathcal{A}'\right)\right)+2 \xi  \mathcal{A} \left(r \mathcal{B}'+\xi  \mathcal{B}^2-\mathcal{B}\right)\right)}{\xi  r^2 \mathcal{A} \mathcal{B}^3}\,,\label{fieldeqsym2}\\
\kappa^2 p_l&=&\frac{1}{2}f+\frac{f_T \left(r \left(r \mathcal{A}' \mathcal{B}'+2 \xi  \mathcal{B}^2 \mathcal{A}'-\mathcal{B} \left(r \mathcal{A}''+3 \mathcal{A}'\right)\right)-\mathcal{A} \left(-r \mathcal{B}'-2 \xi  \mathcal{B}^2+\mathcal{B}^3+\mathcal{B}\right)\right)}{r^2 \mathcal{A} \mathcal{B}^3}-\frac{f_T' \left(r \mathcal{A}'-\xi  \mathcal{A} \mathcal{B}+\mathcal{A}\right)}{r \mathcal{A} \mathcal{B}^2}\nonumber\\
&&+\frac{f_B \left(r \left(r \mathcal{A}' \mathcal{B}'+2 \xi  \mathcal{B}^2 \mathcal{A}'-\mathcal{B} \left(r \mathcal{A}''+4 \mathcal{A}'\right)\right)+2 \mathcal{A} \left(r \mathcal{B}'+\xi  \mathcal{B}^2-\mathcal{B}\right)\right)}{r^2 \mathcal{A} \mathcal{B}^3}+f_B' +\left(\frac{\xi }{r \mathcal{B}}-\frac{\mathcal{B}'}{\mathcal{B}^3}\right)+\frac{f_B''}{\mathcal{B}^2}\,,\label{fieldeqsym3}
\end{eqnarray}
where primes denote differentiation with respect to the radial coordinate, and $\rho,p_r$ and $p_l$ are the energy density, radial pressure and lateral pressure of the fluid. 
In general, the field equations are difficult to solve in an exact form. Some exact solutions for the above system were found in Ref.~\cite{Bahamonde:2019jkf,Golovnev:2021htv}, but the majority of them have a form $\mathcal{B}=\textrm{const.}$, which gives a metric that cannot describe black hole solutions. Moreover, due to the Bianchi identities discussed in section \ref{ssec:BI}, in vacuum, these field equations are not all independent, which simplifies the procedure to find a solution later.

Let us briefly discuss the equations for $f(T)$ gravity in vacuum. For this case, one can manipulate the above system of differential equations by solving~(\ref{fieldeqsym1}) and (\ref{fieldeqsym2}) for $f_{T},f_{T}'$ and replace these expressions in~(\ref{fieldeqsym3}), via Bianchi identities. By doing this, we arrive at the following equation ($f\neq 0$)
\begin{eqnarray}\label{eq:branch1fTB}
0&=&\frac{4 \mathcal{A} \mathcal{B} \left(3 r \mathcal{A}'+2 \mathcal{A}\right)-4 \xi  \mathcal{A} \left(r (\mathcal{B}^2+2) \mathcal{A}'+\mathcal{A} (\mathcal{B}^2+1)\right)}{\mathcal{A} \left(r^2 (\mathcal{A}' \mathcal{B}'-\mathcal{B} \mathcal{A}'')+\mathcal{A} (\mathcal{B}-\mathcal{B}^3)\right)+\xi  \left(r^2 \mathcal{A} \mathcal{A}''+r^2 \mathcal{A}'^2+\mathcal{A}^2 \left(\mathcal{B}^2-1\right)\right)}
\end{eqnarray}
that does not dependent on the form of $f$. This means that the above equation is always true for any form of $f$. In general, the above equation cannot be easily solvable for $\mathcal{A}$ or $\mathcal{B}$. The above equation is the same as the one reported in Ref.~\cite{Golovnev:2021htv} for $\xi=1$. An interesting general feature of $f(T)$ gravity and this branch is the following. By taking a Schwarzschild-like form relating the metric functions as $\mathcal{B}=1/\mathcal{A}$, one can easily solve the above equation giving us that $\mathcal{A}^2=1-2M/r-(\Lambda/3) r^2$. Then, if we replace these expressions into~(\ref{fieldeqsym2}), we find that the unique solution for the system is $f(T)=f_0T-2f_0\Lambda$, which is a trivial case since it is GR plus a cosmological constant. This means that all solutions beyond GR (with the real tetrad) must have a form such that $\mathcal{B}\neq 1/\mathcal{A}$.  

On the other hand, for $f(T,B)$ gravity with $f(T,B)=k_1 T+F(B)$ and after manipulating the field equations~\eqref{fieldeqsym1}-\eqref{fieldeqsym3}, we find that when $k_1=0$ (no GR limit) we must impose the condition $f_B''=F_B''$. Further, for the case $k_1\neq0$, the situation is more complicated that in $f(T)$ gravity since it is not possible to find a similar equation as~\eqref{eq:branch1fTB} that depends on $f$. 

Since it is complicated to find exact solutions beyond $\mathcal{B}=\textrm{const.}$ for the real tetrad, in the following, we will find perturbed solutions. To do this, we consider that the metric is described by Schwarzschild and a small correction related to the modification of GR, namely,
\begin{eqnarray}\label{eq:perturbation}
\mathcal{A}^2(r)&=&1-\frac{2M}{r}+\epsilon\, a(r)\,,\label{eq:schwazr_per_A}\\
\mathcal{B}^2(r)&=&\Big(1-\frac{2M}{r}\Big)^{-1}+\epsilon\, b(r)\label{eq:schwazr_per_B}\,,
\end{eqnarray}
where $\epsilon\ll 1$. Then, one can choose a model, i.e., a form of $f$ such that one has TEGR (or GR) in the background and the form of the correction is also of the order of $\epsilon$. This means that the function $f$ can be written as
\begin{eqnarray}
f(T,B)&=&T+\epsilon\, \tilde{f}(T,B)\,.
\end{eqnarray}
In the following we will consider the following form of $f$
\begin{eqnarray}\label{formf}
\tilde{f}(T,B)=\frac{1}{2}\Big(\alpha T^q+\beta B^m+\gamma B^s T^w \Big)\,,
\end{eqnarray}
where $\alpha,\beta,\gamma,q,m,s$ and $w$ are constants.
The case $\xi=1$ was fully studied in Ref.~\cite{Bahamonde:2020bbc} for different power-law forms of $f(T,B)$ gravity, finding different perturbed solutions around Schwarzschild in vacuum. Furthermore, the power-law $f(T)$ gravity case with $\xi=+1$ was already reported in~\cite{Bahamonde:2019zea} (with $\beta=\gamma=0$) and later in~\cite{Pfeifer:2021njm}, it was found that the squared power-law case (with $q=2$) contains a perturbed black hole solution. On the other hand, for $\xi=-1$, in Ref.~\cite{Ruggiero:2015oka}, the power-law $f(T)$ case was found only when $M=0$. 

Then, we will mainly concentrate in these cases when we analyse the equations around Schwarzschild. As a first step, in the following section, we will analyse the case when we have perturbations around Minkowski. It is important to mention that the limit from the Schwarzschild case to the Minkowski limit is not smooth ($M=0$) since for $\xi=+1$ there are not Minkowski solutions for there are Schwarzschild ones. 

\subsection{Perturbed solutions around Minkowski}
\label{sec:min_per_back}
By plugging~(\ref{formf}) into the field equations~(\ref{fieldeqsym1})-(\ref{fieldeqsym3}) and by assuming the metric functions~(\ref{eq:perturbation}) for the Minkowski case $M=0$, we only find first order perturbed solutions for the case $\xi=-1$ which are explicitly given by
\begin{eqnarray}
a(r)&=&a_0-\frac{a_1}{r}+\beta\frac{   \left(2 m^2-3 m-4\right)2^{3 m-4} }{2 m-3}r^{2-2 m}+\alpha\frac{  2^{3 q-2} }{3-2 q}r^{2-2 q}+\gamma \frac{   \left(2 s^2+s (2 w-3)-4\right)2^{3 s+3 w-4}}{2 s+2 w-3} r^{-2 (s+w-1)}\,,\nonumber \\ \\
b(r)&=&\frac{a_1}{r}+\alpha\frac{   (q-1) (2 q-1)2^{3 q-2}}{2 q-3} r^{2-2 q}+\beta\frac{\left(2 m^3-m^2-3 m+2\right)   8^{m-1} }{2 m-3}r^{2-2 m}\nonumber\\
&&+\gamma\frac{  \left(2 s^3+s^2 (4 w-1)+s \left(2 w^2+3 w-3\right)+4 w^2-6 w+2\right) 2^{3 (s+w-1)}}{2 s+2 w-3} r^{-2 s-2 w+2}\,.
\end{eqnarray}
Here, $a_0$ and $a_1$ are integration constants. The above result generalises the metric found in~\cite{Ruggiero:2015oka} by including the boundary term. The above solution is only valid when $q,m\neq3/2$ and $ s\neq \frac{1}{2}(3-2w)$. There are 6 other solutions which are the possible solutions found from requiring that the denominator in the above cases are zero. For completeness they are explicitly expressed in the appendix~\ref{sec:othersols}. The torsion scalar and boundary term expanded up to first order for the above solution are
\begin{eqnarray}
T&=&\frac{8}{r^2}+\epsilon  \left((-1)^m 8^m\beta   \left(m^2-1\right) r^{-2 m}- 8^q\alpha  (q-1) r^{-2 q}-8^{s+w}\gamma  (s+1)  (s+w-1) r^{-2 s-2 w}\right)\,,\\
B&=&\frac{8}{r^2}+\epsilon  \Big[8^{m-1}\beta   \left(6 m^3-15 m^2-7 m+16\right) r^{-2 m}+ 8^q \alpha \left(q^2-3 q+2\right) r^{-2 q}\nonumber\\
&&+8^{s+w-1}\gamma   \left(6 s^3+3 s^2 (4 w-5)+s \left(6 w^2-7 w-7\right)+8 \left(w^2-3 w+2\right)\right) r^{-2 s-2 w}\Big]\,,
\end{eqnarray}
which  are clearly non-zero even for the case $\alpha=\beta=\gamma=0$. Let us emphasise again here that these solutions appear only when $\xi=-1$, since for the $\xi=1$, there are not perturbed solutions around Minkowski.

\subsection{Perturbed solutions around Schwarzschild}\label{ssec:pertSchwaReal}
Following a similar expansion as the previous section, we will now consider the case $M\neq0$, which corresponds to the case of taking perturbations around Schwarzschild. For this case, one can find solutions for either $\xi=1$ or $\xi=-1$. To integrate out the perturbed equations, we must set the power-law parameters to specific values. Similarly as it was done in~\cite{Bahamonde:2020vpb}, if one expands the function $f$ in a series form of power-laws, the first terms appearing in the expansion would be the ones with $q=m=2$ and $s=w=1$. Therefore, we will concentrate on that case, which gives us the following solutions 
\begin{eqnarray}\label{eq:RealSchw}
a(r)&=&a_0-\frac{a_1}{r}+\frac{1}{6 \xi  r^2 \mu \left(\mu^2-1\right)^2}\Big[3 (2 \beta +\gamma )+3 \xi  \mu^7 (\alpha +13 \beta +7 \gamma )-3 \xi  \mu^5 (15 \alpha +43 \beta +29 \gamma )\nonumber\\
&&+3 \xi  \mu^3 (31 \alpha +51 \beta +41 \gamma -12 (\alpha +\beta +\gamma ) \log (\mu))+3 \xi  \mu (-17 \alpha -21 \beta -19 \gamma +4 (\alpha +\beta +\gamma ) \log (\mu))\nonumber\\
&&+2 \mu^4 (64 \alpha +70 \beta +67 \gamma )-9 (2 \beta +\gamma ) \mu^8+12 (2 \beta +\gamma ) \mu^6-12 (2 \beta +\gamma ) \mu^2\Big]\,,\\
b(r)&=&a_0\frac{\mu ^2-1}{\mu^4}+\frac{a_1}{r \mu^4}+\frac{1}{6 \xi ^5 r^2 \mu^5 \left(\mu^2-1\right)}\Big[6 (2 \beta +\gamma )+3 \xi  \mu^5 (25 \alpha +37 \beta +31 \gamma )\nonumber\\
&&-12 \xi  \mu^3 (\alpha +3 \beta +2 \gamma )+3 \xi  \mu (-21 \alpha -25 \beta -23 \gamma +4 (\alpha +\beta +\gamma ) \log (\mu))-12 \mu^6 (2 \alpha +4 \beta +3 \gamma )\nonumber\\
&&-2 \mu^4 (32 \alpha +26 \beta +29 \gamma )+24 \mu^2 (\alpha +\beta +\gamma )\Big]\,,
\end{eqnarray}
where $\mu=\sqrt{1-2M/r}$ and $a_0,a_1$ are integration constants. These solutions were found in~\cite{Bahamonde:2020bbc} for $\xi=1$ and the case $\xi=-1$ has not been reported yet. 

To ensure that the perturbed spacetime geometry has a putative horizon and can be interpreted as black hole spacetime which perturbs Schwarzschild geometry, we follow the procedure which was applied in \cite{Pfeifer:2021njm} to a subclass of the solutions presented here. 

We fix one of the constants of integration at the putative horizon $r=r_h$ where $\mathcal{A}(r_h)=0$, by demanding that the product of $\mathcal{A}$ and $\mathcal{B}$ stays finite. In the perturbative approach this amounts to
\begin{align}
    \lim_{r\to r_h} (\mathcal{A}(r)\mathcal{B}(r))^2 = 1+ \epsilon \mathfrak{h}\,,\label{horizon}
\end{align}
where $\mathfrak{h}$ is a constant. To fix the other constant we apply an asymptotic expansion for $r\to\infty$ to $\mathcal{A}(r)$ and $\mathcal{B}(r)$ and demand that $\mathcal{B}(r)$ behaves like the Schwarzschild solution to order $1/r$.

It turns out that the equation~\eqref{horizon} only depends on $a_0$, which is determined to be
\begin{align}\label{eq:a0real}
    a_0 = \frac{1}{24 M^2 \mu_h^3 \xi}
    \bigg(& 3 (2 \beta +\gamma )+3 \mu_h^8 (8 \alpha +22 \beta +15 \gamma )+10 \mu_h^6 (4 \alpha -2 \beta +\gamma )-216 \mu_h^4 (\alpha +\beta +\gamma )+6 \mu_h^2 (4 \alpha +6 \beta +5 \gamma ) \nonumber\\
    &+\xi  \big(6 \mu_h^3 \left(-7 \alpha -19 \beta -13 \gamma +4 \mathfrak{h}  M^2+4 \log (\mu_h) (\alpha +\beta +\gamma )\right)+6 \mu_h^7 (-13 \alpha -25 \beta -19 \gamma )\nonumber\\
    &+12 \mu_h^5 (11 \alpha +23 \beta +17 \gamma )-12 \mu_h (\alpha +\beta +\gamma ) \big)\bigg)\,,
\end{align}
where $\mu_h = \sqrt{1-\frac{2 M}{r_h}}$.

By studying the asymptotic expansion of (\ref{eq:RealSchw}) we find,
\begin{eqnarray}
    \mathcal{A}^2(r)
    &=& 1 -\frac{2 M}{r} + \epsilon \left(a_0 + \frac{a_1}{r} -\frac{16 (3 \frac{M}{r}-1) (\alpha +\beta +\gamma )}{3 M^2 \xi }\right) + \mathcal{O}\Big(\frac{1}{r^2}\Big)\,,\label{eq:Schwar_per_sol_A}\\
    \mathcal{B}^2(r)
    &=& 1 + \frac{2 M}{r} + \frac{\epsilon}{r} \left(-2 a_0 M+a_1+\frac{16 ( \alpha +\beta +\gamma )}{3 M \xi }\right) + \mathcal{O}\Big(\frac{1}{r^2}\Big)\,.\label{eq:Schwar_per_sol_B}
\end{eqnarray}
The desired fall off behaviour for $\mathcal{B}(r)$ yields 
\begin{align}\label{eq:a1real}
    a_1 = 2 a_0 M-\frac{16 (\alpha +\beta +\gamma )}{3 M \xi }\,,
\end{align}
which makes 
\begin{equation}
    \mathcal{A}^2(r) = \Big(1-\frac{2M}{r}\Big)\Big(1+\epsilon\,  \left[\frac{16}{3 M^2 \xi }  (\alpha +\beta +\gamma )+  a_0 \right]\Big) + \mathcal{O}\Big(\frac{1}{r^2}\Big)
\end{equation}
for $r\to\infty$.

Having fixed the integration constant by the horizon condition and by a desired fall of behaviour for large $r$, we cannot impose more constraints to obtain a desired behaviour of the solutions in the $M\to0$ limit. We observe that in the $M\to 0$ limit we find
\begin{align}
    a(r) &\to \frac{4 (\xi -1) (6 \alpha +8 \beta +7 \gamma )+\mathfrak{h}  r^2}{\xi  r^2}\,,\qquad
    b(r) \to -\frac{4 (\xi -1) (6 \alpha +8 \beta +7 \gamma )}{\xi  r^2}\,.
\end{align}
Thus, these functions always vanish for $\xi=1$, and so these solutions have a smooth consistent Minkowski spacetime limit. For $\xi=-1$, for generic $\alpha, \beta, \gamma$, $\mathcal{A}^2(r)$ and $\mathcal{B}^2(r)$ the solutions do not converge to Minkowski spacetime for $M\to 0$, but have first order corrections which go with $r^{-2}$. Observationally, this manifests itself in the Shapiro delay, which is nontrivial for $M\to0$ for $\xi=-1$ (see Sec.~\ref{sec:phenomenology}). The only choice one has, if one wants that the $\xi=-1$ tetrad has a ``good" $M\to 0$ Minkowski limit, is to choose $6 \alpha +8 \beta +7 \gamma=0$ and $\mathfrak{h} =0$, otherwise the torsion sources curvature.

\section{Symmetric field equations and solutions for the complex tetrad}\label{sec:complex_tetrad}
This section will be devoted to studying the complex tetrad~(\ref{tetrad2}) and obtain solutions. By using this tetrad in the $f(T,B)$ gravity equations~(\ref{fieldeq}), we find that the remaining field equations (symmetric field equations) for the complex tetrad become
\begin{eqnarray}
\kappa^2\rho&=&-\frac{1}{2} f+\frac{2 f_T \left(r \mathcal{B} \mathcal{A}'+\mathcal{A} \left(\mathcal{B}-r \mathcal{B}'\right)\right)}{r^2 \mathcal{A} \mathcal{B}^3}+\frac{f_B \left(r \left(\mathcal{B} \left(r \mathcal{A}''+4 \mathcal{A}'\right)-r \mathcal{A}' \mathcal{B}'\right)+2 \mathcal{A} \left(\mathcal{B}-r \mathcal{B}'\right)\right)}{r^2 \mathcal{A} \mathcal{B}^3}\nonumber\\
&&+\frac{\mathcal{B}' f_B'}{\mathcal{B}^3}-\frac{f_B''}{\mathcal{B}^2}+\frac{2 f_T'}{r \mathcal{B}^2}\,,\label{img1}\\
\kappa^2 p_r&=&\frac{1}{2} f+\frac{\left(r \mathcal{A}'+2 \mathcal{A}\right) f_B'}{r \mathcal{A} \mathcal{B}^2}-\frac{2 f_T \left(2 r \mathcal{A}'+\mathcal{A}\right)}{r^2 \mathcal{A} \mathcal{B}^2}+\frac{f_B \left(r \left(r \mathcal{A}' \mathcal{B}'-\mathcal{B} \left(r \mathcal{A}''+4 \mathcal{A}'\right)\right)-2 \mathcal{A} \left(\mathcal{B}-r \mathcal{B}'\right)\right)}{r^2 \mathcal{A} \mathcal{B}^3}
\,,\label{img2}\\
\kappa^2 p_l &=&\frac{1}{2} f-\frac{\left(r \mathcal{A}'+\mathcal{A}\right) f_T'}{r \mathcal{A} \mathcal{B}^2}+\frac{f_B \left(r \left(r \mathcal{A}' \mathcal{B}'-\mathcal{B} \left(r \mathcal{A}''+4 \mathcal{A}'\right)\right)-2 \mathcal{A} \left(\mathcal{B}-r \mathcal{B}'\right)\right)}{r^2 \mathcal{A} \mathcal{B}^3}\nonumber\\
&&+\frac{f_T \left(r \left(r \mathcal{A}' \mathcal{B}'-\mathcal{B} \left(r \mathcal{A}''+3 \mathcal{A}'\right)\right)-\mathcal{A} \left(-r \mathcal{B}'+\mathcal{B}^3+\mathcal{B}\right)\right)}{r^2 \mathcal{A} \mathcal{B}^3}-\frac{\mathcal{B}' f_B'}{\mathcal{B}^3}+\frac{f_B''}{\mathcal{B}^2}\,.\label{img3}
\end{eqnarray}
In the following we will use the same techniques as in the previous tetrad to obtain solutions around Minkowski and  Schwarzschild, and then, we will solve these equations in an exact form (for some specific cases). 

\subsection{Perturbed solutions around Minkowski}
\label{sec_com_mn_per}
Around Minkowski ($M=0$) and assuming that the function has the form~(\ref{formf}) we find the following solution up to first order in $\epsilon$:
\begin{eqnarray}
a(r)&=&a_0-\frac{a_1}{r}+\alpha \frac{ 4^{q-1} }{3-2 q}r^{2-2 q}+\beta \frac{  \left(2 m^2-3 m-2\right) 2^{2 m-3}}{2 m-3}r^{2-2 m}+\gamma\frac{   \left(2 s^2+s (2 w-3)-2\right)2^{2 s+2 w-3}}{2 s+2 w-3} r^{-2 s-2 w+2}\,,\nonumber \\ \\
b(r)&=&\frac{a_1}{r}+\alpha \frac{  (q-1) (2 q-1)4^{q-1}}{2 q-3} r^{2-2 q}+\beta\frac{ 4^{m-1} (m-1)^2  (2 m+1) }{2 m-3}r^{2-2 m}\nonumber\\
&&+\gamma\frac{   \left(2 s^3+s^2 (4 w-3)+s w (2 w-1)+2 w^2-3 w+1\right) 2^{2 (s+w-1)}}{2 s+2 w-3}r^{-2 s-2 w+2}\,,
\end{eqnarray}
where $a_0$ and $a_1$ are constants. As it happened in the previous section, there are other possible solutions for  particular cases containing logarithmic terms and for completeness they are written in appendix~\ref{sec:othersols2}.

\subsection{Perturbed solutions around Schwarzschild}\label{ssec:pertSchwaIm}
In general, for $M\neq0 $, we cannot find an analytic expression for perturbative solutions of the model~(\ref{formf}) for an arbitrary parameter $q$. This is not the case when one decouples the boundary term by choosing $\beta=\gamma=0$, for which it is possible to analytically integrate the equations for any power-law parameter $q$ which is related to the torsion scalar part. This property suggests that the dynamics of the field equations arising from the complex tetrad would be different to the ones arising from the real tetrad.
Recall that  for the real tetrad, one must impose the value of the torsion scalar power-law parameter to obtain perturbed  solutions (see for example the solution with $q=2$ presented in~\eqref{eq:RealSchw}). 

If one wants to include the boundary term contributions, we can now choose $m=2$ and $s=w=1$ in~(\ref{formf}) in order to have the first order terms that would appear in a Taylor series for the boundary contribution in $f$. For this case, after expanding the equations~(\ref{img1})-(\ref{img3}) up to first order in $\epsilon$, we find the following class of solutions
\begin{eqnarray}\label{eq:CompSchw}
a(r)&=&a_0-\frac{a_1}{r}+\alpha\frac{  4^{q-1} }{3-2 q}r^{2-2 q}-\frac{6 M (2 \beta +\gamma )}{r^3}-\frac{2 \gamma }{r^2}\,,\\
b(r)&=&\frac{1}{(r-2 M)^2}\Big[a_1 r-2 a_0 M r-2^{2 q-1}M  q \alpha   r^{3-2 q}+\alpha\frac{  (q-1) (2 q-1) 4^{q-1}}{2 q-3} r^{4-2 q}\nonumber\\
&&+4 (5 \beta +4 \gamma )-\frac{2 M (14 \beta +11 \gamma )}{r}\Big]\,,
\end{eqnarray}
where $a_0,a_1$ are integration constants.

In the same way as we fixed the constants of integration for the real tetrad case in Section \ref{ssec:pertSchwaReal} in equations \eqref{eq:a0real} and \eqref{eq:a1real}, we now also fix them here.

First we demand the horizon condition $\lim_{r\to r_h}(\mathcal{A}(r)\mathcal{B}(r))^2 = 1 + \epsilon \mathfrak{h}$, which only depends on $a_0$, and find that
\begin{align}
    a_0 = \alpha  q \left(-M^{2-2 q}\right) \left(1-\mu_h^2\right)^{2 (q-1)}-\frac{\left(\mu_h^2-1\right)^2 (10 \beta +7 \gamma )}{2 M^2} + \mathfrak{h}\,,
\end{align}
where $\mu_h = \sqrt{1-\frac{2 M}{r_h}}$ and it depends on the value of $q$ if the $q$-dependent term must be taken into account in this order of the expansion, or not.
Second we study the asymptotic expansion of (\ref{eq:CompSchw}) for $r\to\infty$ and find,
\begin{eqnarray}
    \mathcal{A}^2(r) 
    &=& 1 - \frac{2 M}{r} + \epsilon  \left(a_0-  \tfrac{a_1}{r}+ \frac{ 4^{q-1}\alpha  }{3-2 q}\right)\frac{1}{r^{2 q-2}} + \mathcal{O}\Big(\frac{1}{r^2}\Big)\\
    \mathcal{B}^2(r) 
    &=& 1 + \frac{2 M}{r} 
    + \frac{\epsilon}{r} \left( a_1 - 2 M a_0 \right)\nonumber \\
    &+& \epsilon \frac{ 4^{q-1}\alpha }{2 q-3} \left(4 M^2 (q (2 q-3)+3 ) \frac{1}{r^{2q}}+2 M (q (2 q-3)+2) \frac{1}{r^{2q-1}}+(q-1) (2 q-1) \frac{1}{r^{2q-2}}\right) + \mathcal{O}\Big(\frac{1}{r^2}\Big)\,.
\end{eqnarray}
Again the precise behaviour of the components depends on the choice of the parameter $q$.
The desired fall off behaviour for $\mathcal{B}(r)=1 + \tfrac{2 M}{r}+ \mathcal{O}(\tfrac{1}{r^2})$ can only be achieved for $q\geq 1$. For $q\in\mathbb{N}\geq 1$ we obtain two cases (the different possibilities for $q\in\mathbb{R}^+\geq 1$ have to be investigated separately)
\begin{align}\label{eq:a1ofa0cmplx}
    a_1|_{q=1} = 2 M (a_0 + \alpha),\quad   a_1|_{q>1} = 2 M a_0\,.
\end{align}
In these cases we find
\begin{align}
     \mathcal{A}^2(r)|_{q=1} = \left(1 - \frac{2M}{r}\right)(1+\epsilon (a_0 + \alpha)) + \mathcal{O}\Big(\frac{1}{r^2}\Big),\quad \mathcal{A}^2(r)|_{q>1} = \left(1 - \frac{2M}{r}\right)(1+\epsilon a_0) + \mathcal{O}\Big(\frac{1}{r^2}\Big)\,.
\end{align}
Similar, as we discussed in Sec. \ref{ssec:pertSchwaReal} for the real tetrad with $\xi=-1$, the $M\to0$ limit does always yield Minkowski spacetime geometry but the corrections of order $\frac{1}{r^2}$ yield non trivial curvature terms, which manifest themselves for example in a non-trivial Shapiro delay, see Sec.~\ref{sec:phenomenology}.

\subsection{Exact solutions for \texorpdfstring{$f(T)$}{} gravity}
\label{sec:com_f_T_sol}
This section will be devoted to find exact spherically symmetric solutions for $f(T)$ gravity associated with the complex tetrad. One first notices that the system~(\ref{img1})-(\ref{img3}) is less involved than the one described by the real tetrad~(\ref{fieldeqsym1})-(\ref{fieldeqsym3}). 
When considering the form of the torsion scalar given by, Eq.~(\ref{eq:torsionscalarcomplex}), it is found an equation for $f$ and $T$, which can be solved if finding $T$ as a function of $r$, and provided a functional form for $f$ 
\begin{eqnarray}
\left(\frac{2}{r^2}-T\right) f'(T)+\frac{f(T)}{2}=0\,.
\label{fTeq}
\end{eqnarray}
Similarly as we did before, and using Bianchi identities, one notices that for the $f(T)$ gravity case, we can use~(\ref{img1})-(\ref{img2}) to eliminate all the $f$ dependence by solving for $f_T,f_T'$ and replacing these expressions in~(\ref{img3}). It turns out that after this manipulation one obtains the following equation
\begin{eqnarray}
-f\frac{\left(r^2 \mathcal{A} \mathcal{A}''+r^2 \mathcal{A}'^2+\mathcal{A}^2 \left(\mathcal{B}^2-1\right)\right) }{4 \mathcal{A} \left(2 r \mathcal{A}'+\mathcal{A}\right)}=0\,,
\end{eqnarray}
that it is much simpler than~(\ref{eq:branch1fTB}) since it can be easily solved for $\mathcal{B}$, giving us
\begin{equation}
\label{Bcomplex}
\mathcal{B}  = \pm \frac{\sqrt{-r^2 \mathcal{A} \mathcal{A}''-r^2 \mathcal{A}'^2+\mathcal{A}^2}}{\mathcal{A}}\,.
\end{equation}
The above equation tells us that the form of the metric function $g_{rr}$ will always have the same form independently  of the form of $f$. 

The torsion scalar $T$ given by Eq.~(\ref{eq:torsionscalarcomplex}) for this case yields 
\begin{equation}\label{Tscalar}
T=\dfrac{2 (-2 \mathcal{A}^2 + r^2 \mathcal{A}'^2 +  r \mathcal{A}(-2 \mathcal{A}' + r \mathcal{A}''))}{(-r^2 \mathcal{A}^2 +  r^4 \mathcal{A}'^2 + r^4 \mathcal{A} \mathcal{A}'')}\,.
\end{equation}
Now, the equation \eqref{fTeq} cannot be solved unless we  assume either $\mathcal{A}$ or the form of $f(T)$ since we would need to replace either $T=T(r)$ or $r=r(T)$. In the following, we will explore both possible ways of obtaining exact solutions for the complex tetrad.

\subsubsection{Choosing the metric function \texorpdfstring{$\mathcal{A}(r)$}{}}
In this section we will assume a form for $\mathcal{A}(r)$ and then find the theory $f(T)$ that reproduces that specific form of the metric. This procedure is similar to the reconstruction method employed in cosmology when the scale factor is assumed and then the form of the Lagrangian is found.

The first conclusion that we can obtain is that after assuming a Schwarzschild-like form $\mathcal{B}=1/\mathcal{A}$, the unique solution for (\ref{Bcomplex}) is $\mathcal{A}^2=1-2M/r+c r^2$, which is a Schwarzschild de-Sitter solution. By plugging this solution in~(\ref{fTeq}) we find that   $f(T)=T+\textrm{const.}$. This means that, as we pointed out in the previous section for the other tetrad, the unique solution behaving as $\mathcal{B}=1/\mathcal{A}$ is the Schwarzschild-De Sitter with the trivial GR$+\Lambda$ Lagrangian. This means that for the two possible branches of tetrads that satisfy the antisymmetric field equations in spherical symmetry (which satisfy that the metric and teleparallel connection respect spherical symmetries), we must have that $g_{tt}\neq - g_{rr}$ to have spherically symmetric solutions beyond GR. This important remark will be used later in the conclusions (see \ref{sec:conclusion}) to elaborate a theorem for $f(T)$ gravity.

The simplest choice to obtain exact solutions is by imposing that the metric function $\mathcal{B}(r)=\text{const}:=b$. By replacing this ansatz in (\ref{Bcomplex}), the following equation is obtained for $\mathcal{A}$
\begin{equation}
r^2 \dfrac{\mathcal{A}''}{\mathcal{A}} + r^2 \dfrac{\mathcal{A}'^2}{\mathcal{A}^2} + b-1 = 0\,.
\end{equation}
This equation can be analytically solved yielding
\begin{equation}
 \mathcal{A}(r)=a_2 r^{\frac{1}{4}-\frac{1}{4} \sqrt{9-8 b}} \sqrt{a_1+r^{\sqrt{9-8 b}}}\,,
\end{equation} where $a_1,a_{2}$ are integration constants. When $a_1=0$ we can easily solve~\eqref{fTeq} to obtain $f(r)$ in terms of $r$ and then we can solve $r=r(T)$ and get that the form $f(T)$ which reproduces the above metric function with $a_1=0$ becomes
\begin{equation}
    f(T)=F_0\left(\sqrt{9-8 b}+3\right) b^u \left(2 b+\sqrt{9-8 b}+3\right)^{-u} T^u\,,\quad u=\frac{1}{8} \left(7-\sqrt{9-8 b}\right)\,.
\end{equation}
Even though this is an exact spherically symmetric solution, it cannot describe a black hole geometry since by construction we imposed $g_{rr}=-\textrm{const.}$

In order to find physically meaningful solutions, we can assume a more general ansatz for the metric function $\mathcal{A}(r)$ expressed by 
\begin{equation}
    \mathcal{A}^2=1-2M/r+Q/r^q\,,
    \label{Acharg}
\end{equation} 
which contains a Schwarzschild like term plus an arbitrary power-law form depending on a parameter $q$. Recall that the form of $\mathcal{B}(r)$ will be given by~\eqref{Bcomplex}, so that, by imposing the above form of $\mathcal{A}(r)$, the metric is already determined.
To find out solutions, we first replace \eqref{Acharg} in~\eqref{Tscalar}, and one can rewrite in~\eqref{fTeq} the derivative as
$f'(T)=f'(r)(dT/dr)^{-1}$, yielding the following
\begin{equation}
  f'(r)  \frac{r  \left(\left(q^3-3 q+2\right) Q^2-\left(q^2+3 q-4\right) Q r^q+2 r^{2 q}\right)}{(q-1)^2 \left(q^2+6 q+8\right) Q^2+\left(q^3-7 q^2-10 q+16\right) Q r^q+8 r^{2 q}}+\frac{f(r)}{2}=0\,.\label{eq:com_f_T_sol1}
\end{equation}
The above equation can be solved analytically for $f(r)$ for $q\neq-4$
\begin{equation}\label{f1}
    f(r)=f_0\frac{2 \sqrt{2}  r^{-\frac{q}{2}-2} \left(r^q+Q-q Q\right)^{7/4}}{\left(2 r^q-\left(q^2+q-2\right) Q\right)^{5/4}}\,,
\end{equation}
while for $q=4$ we obtain
\begin{equation}
f(r)= f_0\frac{ \left(1+5 Q r^4\right)}{r^2 \sqrt{1-5 Q r^4}}\,.\label{eq:com_f_T_sol2}
\end{equation}
Next, we need to invert $f(r)=f(T)$. This can be done by replacing our ansatz in the torsion scalar  (see Eq.~\eqref{Tscalar}), giving
\begin{equation}
    T=T(r)=\frac{8 r^q-2 \left(q^2+3 q-4\right) Q}{r^2 \left(2 r^q-\left(q^2+q-2\right) Q\right)},
\end{equation}
and proceeding to invert in order to find $r=r(T)$.

One interesting solution is the one by choosing $q=2$. In this case, the torsion scalar is $T=(4 r^2-6 Q)/(r^4-2 Q r^2)$, and then one can solve this relationship for $r=r(T)$,
\begin{equation}
r_{\pm} =  \sqrt{Q + \dfrac{2}{T} \pm \dfrac{\sqrt{4-2QT +Q^2 T^2} }{T} }\,,\label{rplusminus}
\end{equation}
giving us that the form of $f(r)$ given in~\eqref{f1} can be expressed as $f(T)$ as follows 
\begin{eqnarray}
f(T)=4 f_0 \, \frac{\left(2\pm\sqrt{Q^2 T^2-2 Q T+4}\right)}{\left(Q T+2\pm\sqrt{Q^2 T^2-2 Q T+4}\right) \sqrt{8-2 Q T\pm 4 \sqrt{Q^2 T^2-2Q T+4}}}\,,
\end{eqnarray} 
where for the plus(minus) sign we must have $r^2\geq 2Q$($r^2\leq 2Q$). By assuming  $Q\ll 1$ and expanding up to first order expansions we get $f(T)=f_0 (T-\frac{1}{8}QT^2)$, which is a squared power-law $f(T)$ gravity.

It is important to mention that in this particular case $q=2$, the metric is
\begin{equation}\label{sol1}
      ds^2=\Big(1-\frac{2M}{r}+\frac{Q}{r^2}\Big)dt^2-\Big(\frac{2 Mr-Q-r^2}{2 Q-r^2}\Big)^{-1}dr^2-r^2d\Omega^2\,,
\end{equation}
which behaves similarly to the Reissner–Nordstr\"om one but with $g_{rr}\neq -1/g_{tt}$. Even though $g_{rr}\neq -1/g_{tt}$, it is easy to see that this metric is an exact black hole solution with two horizons $r_{h,\pm}=M\pm\sqrt{M^2+Q}$, since at the horizons we have that $g_{tt}|_{r=r_{h,\pm}}=-1/g_{rr}|_{r=r_{h,\pm}}=0$, while $\textrm{det}(g_{\mu\nu})|_{r=r_{h,\pm}}$ is regular. To the best of our knowledge, this is the first non-trivial exact black hole solution in modified teleparallel gravity. Recall that since $Q$ is a constant that it is not related to the electromagnetic charge, our black hole solution can have $Q<0$ as well. If we assume that $T<0$ and $Q=0$, the above form of $f(T)\propto T$ as expected since one recovers the Schwarzschild metric. It is also worth noticing that in the context of metric-affine gravity with curvature, nonmetricity and torsion (propagating), a Reissner–Nordstr\"om black hole solution was also found where if one ignores the nonmetricity contribution, $Q$ is related to the spin charge~\cite{Bahamonde:2021akc,Bahamonde:2020fnq}.

\subsubsection{Choosing the form of \texorpdfstring{$f(T)$}{}}
Another way to solve the system of equations is by assuming a form of $f(T)$. This method seems more physically interesting, since one can propose a specific theory in the functional form of $f(T)$ and the output will be a differential equation for $\mathcal{A}(r)$.

In order to illustrate this, let us take a power-law form $f(T)=k_1 T+(1/2)\alpha T^p$.  By replacing this form in Eq.~(\ref{fTeq}), we find the following differential equation for $p=2$ 
\begin{equation}
   0 =k_1 r^3 u'(r)-3 \left(5 \alpha +k_1 r^2\right)+\frac{\left(\alpha +k_1 r^2\right) \left(r u'(r)-2\right)^2}{u(r)^2}-u(r) \left(16 \alpha +k_1 r^2\right)-\frac{2 \alpha  \left(r u'(r)-2\right)}{u(r)}-4 \alpha  u(r)^2\,,
\end{equation}
where we introduced $u(r)=\mathcal{A}/(r\mathcal{A}')$. This equation is not easily solvable for $k_1\neq0$ but its form is simpler without introducing the variable $u$. When $k_1=0$ (that is, no GR background), one can solve the equations for an even power-law $p=2,4,6,..$. Explicitly, for this case we find two different solutions, the first one given by
\begin{eqnarray}
\mathcal{A}(r)^2&=&\frac{c_1}{r^{4 (p-1)}}+\frac{c_2}{r}\,,
\end{eqnarray}
while the second is
\begin{eqnarray}
\mathcal{A}(r)^2&=&c_2 r^4+\frac{c_1}{r}\,.
\end{eqnarray}
These two exact solutions are not asymptotically flat nor can represent a black hole solution. Moreover, they give a torsion scalar $T=0$, and at the level of the equations of motion they impose $f(0)=f_T(0)=0$, which represents a trivial solution of the equations of motion with switched off gravity. When $k_1\neq0$ (GR background), the equations become much more involved and it is hard to obtain analytic solutions. 

A more interesting option is to take a Born-Infeld $f(T)$ gravity whose form is
\begin{equation}
    f(T)=\lambda\Big(\sqrt{1+\frac{2T}{\lambda}}-1\Big)\,,
\end{equation}
with $\lambda$ being the so-called Born-Infeld parameter. It is easy to notice that when $T/\lambda\ll 1$, one obtains $f(T)=T-T^2/(2\lambda)+\mathcal{O}(1/\lambda^2)$. If we replace this functional form in~\eqref{fTeq} and use both \eqref{Bcomplex} and the form of the torsion scalar~\eqref{eq:torsionscalarcomplex}, we arrive at the following exact solution
\begin{equation}
    ds^2=\frac{a_1^2 }{r}\Big[\sqrt{\lambda } (a_0 \lambda +r)-2 \tan ^{-1}\left(\frac{\sqrt{\lambda } r}{2}\right)\Big]dt^2-\frac{\lambda ^{5/2} r^5}{(4 + r^2 \lambda)^2}\Big[\sqrt{\lambda } (a_0 \lambda +r)-2 \tan ^{-1}\left(\frac{\sqrt{\lambda } r}{2}\right)\Big]^{-1}dr^2-r^2d\Omega^2\,,\label{eq:com_f_T_metric}
\end{equation}
where $a_0,a_1$ are integration constants. This is a nontrivial solution of the $f(T)$ equations of motion for our complex tetrad, since this time the torsion scalar $T = (8+4r^2\lambda)/(r^4\lambda)$ is not constant and in particular not zero. Most interestingly, it is independent of the integration constants $a_0$ and $a_1$. 

\begin{figure}[H]
    \centering
    \includegraphics[scale=1]{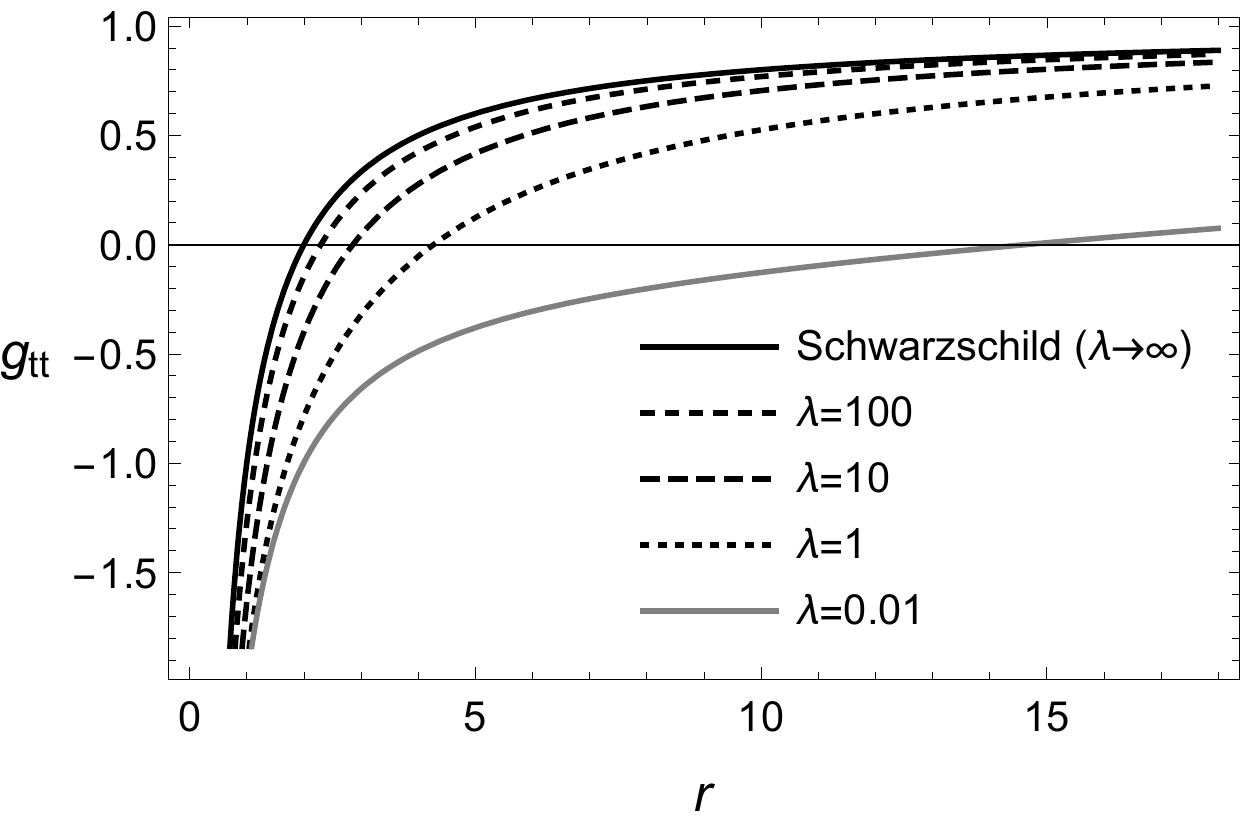}
    \caption{Metric component $g_{tt}$ versus $r$ for the solution~\eqref{eq:com_f_T_metric} with different values of $\lambda$. We have set $a_0=-2M/\lambda$, $a_1^2=1/\sqrt{\lambda}$ to ensure asymptotically flatness and a smooth transition from Schwarzschild. We have also set $M=1$.}
    \label{fig1}
\end{figure}

In order to obtain an asymptotically flat spacetime we require that  $a_1^2=1/\sqrt{\lambda}$. If one wants to show that the above metric can describe a black hole, one would need to solve $g_{tt}=0$ and check that the determinant of the metric is non-singular. Assuming that $\lambda>0$, one can easily see that the determinant is always non-singular. It is given by $\sqrt{-g}=\textrm{det}(h^A{}_\mu)\equiv h=(\lambda  r^4 \sin \vartheta)/(\lambda  r^2+4)$. Solving $g_{tt}=0$ cannot be done analytically easily. In Fig.~\ref{fig1}, we show the behaviour of $g_{tt}(r)$ for different values of $\lambda$. We can notice that there is one horizon showing that the above metric can describe a black hole. Further, the smaller $\lambda$ is, the larger is its influence and the horizon of the black hole is shifted to a greater value compared to Schwarzschild geometry.

Having fixed one of the integration constants such that the metric is asymptotically flat and expanding to powers of  $1/\lambda$, we find that, in order to have a smooth transition from the Schwarzschild solution of TEGR ($\lambda\rightarrow \infty$) one needs to set $a_0=-2M/\lambda$. Doing so, the expansion of the metric up to $\mathcal{O}(1/\lambda^2)$ becomes
\begin{equation}
      ds^2=\Big[1-\frac{2 M}{r}+\frac{4}{\lambda  r^2}-\frac{\pi }{\sqrt{\lambda } r}\Big]dt^2-\Big[1-\frac{2 M}{r}-\frac{16 M}{\lambda  r^3}+\frac{12}{\lambda  r^2}-\frac{\pi }{\sqrt{\lambda } r}\Big]^{-1}dr^2-r^2d\Omega^2+\mathcal{O}(1/\lambda^2)\,.
\end{equation}

In this approximation we can calculate an explicit expression for the horizon radius to leading order in $1/\lambda$, and then solve order by order. The ansatz $r_h = 2M +  r_1/\sqrt{\lambda} + r_2/\lambda $ yields
 \begin{eqnarray}
 r_{h}= 2M + \frac{\pi}{\sqrt{\lambda}} - \frac{2M}{\lambda} +\mathcal{O}(1/\lambda^2)\,.
 \end{eqnarray} 
Thus, we demonstrated how the solution \eqref{eq:com_f_T_metric} is a generalisation of a Schwarzschild solution.

It would be of interest to explore if a match of our exact exterior black hole solutions with interior solutions for the real tetrad already found in the literature \cite{Bohmer:2019vff,Boehmer:2020hkn} is possible. Such matching could be studied in the lines of the formalism for matching tetrad solutions proposed in  \cite{Fiorini:2021mps}.

\subsection{Exact solutions for \texorpdfstring{$f(T,B)=k_1T+F(B)$}{} gravity}

A similar analysis can be done for $f(T,B)=k_1 T+F(B)$ gravity. For this case, one can again manipulate Eqs~\eqref{img1}-\eqref{img3} to obtain an independent $F(B)$ equation which reads
\begin{eqnarray}
0&=&-\frac{k_1 \left(r \left(\mathcal{B} \left(r \mathcal{A}''+\mathcal{A}'\right)-r \mathcal{A}' \mathcal{B}'\right)+\mathcal{A} \left(r \mathcal{B}'+\mathcal{B}^3-\mathcal{B}\right)\right)}{r^2 A \mathcal{B}^3}\,.
\end{eqnarray}
The situation becomes more complicated for $f(T,B)$ gravity since even for this complex tetrad, one cannot easily solved the above equation (as it happens in $f(T)$). To solve this equation, let us assume that $\mathcal{B}=B_0 r^k/\mathcal{A}$, which gives us
\begin{equation}
    \mathcal{A}(r)=\frac{r^{\frac{2-7 k}{8 (k-1)}} \sqrt{2 \left(k^2-1\right)K_0 r^{\frac{5 (3 k-2)}{4 (k-1)}}-(k-3) r^{\frac{4 k^2-k+2}{4 (k-1)}} \left(B_0^2 r^{k+1}-k^2 K_1+K_1\right)}}{\sqrt{k^3-3 k^2-k+3}}\,,\quad k\neq \pm 1,3\,.\label{eq:com_f_B_solA}
\end{equation}
The form of $\mathcal{A}(r)$ for $k=\pm 1,3$  by logarithmic terms and for some of them, it is possible to solve the remaining field equation to get $F(B)$, but all of those solutions are not asymptotically flat and due to the logarithmic terms, they might not behave as black holes. 

One notices that the unique asymptotically flat metric is when $K_0=k=0$. Using this solution in the remaining field equation~\eqref{img2}, we find that for $k=0$ ($\mathcal{B}=1/\mathcal{A}$), the solution behaves as Schwarzschild-de Sitter: ($B_0=1$)
\begin{equation}
    ds^2=\Big(1-\frac{2M}{r}-(\Lambda+c_0 M)r^2\Big)dt^2-\Big(1-\frac{2M}{r}-(\Lambda+c_0 M)r^2\Big)^{-1}-r^2d\Omega^2\,,\label{eq:com_f_B_sol1}
\end{equation}
 with $f(T,B)$ being given by
\begin{equation}
    f(T,B)=\frac{T}{2}-\frac{8c_0}{3 \sqrt{B+18 (Mc_0+\Lambda )}}-3\Lambda\,,
\end{equation}
where for simplicity we have set $k_1=1/2$.
This solution has an effective cosmological constant that depends on the boundary term contribution $\Lambda_{\rm eff}=\Lambda+c_0 M$. Thus, the unique solution for $f(T,B)=k_1+F(B)$ with $\mathcal{A}=1/\mathcal{B}$ is a Schwarzschild de-Sitter metric with an effective cosmological constant whose contribution is related to the boundary term.

For other values of $k\neq 0$, it is also possible to solve~\eqref{img2} for $F=F(r)$ but only for some of them, it is possible to invert $r=r(B)$ to find out an explicit form of $F(B)$ in terms of the boundary term. All of those solutions are not asymptotically flat. For example for $k=2$ one finds that the form of $f$ behaves as (with $B_0=k_1=1$)
\begin{eqnarray}
f(T,B)=T-\frac{4  \left(\sqrt{4-3 B K_0}-2\right)}{15 K_0}+\frac{1}{25} B  \left[10 \log \left(-\frac{2 \left(\sqrt{4-3 B K_0}+2\right)}{B}\right)-1\right]\,,
\end{eqnarray}
while the metric becomes
\begin{equation}\label{sol3}
    ds^2=\Big(K_1 r-2 K_0 r^2-\frac{r^4}{3}\Big)dt^2-r^2\Big(K_1 r-2 K_0 r^2-\frac{r^4}{3}\Big)^{-1}dr^2-r^2d\Omega^2\,,
\end{equation}
which can represent a non-asymptotically flat black hole solution in $f(T,B)$ gravity with one horizon.

\section{Particle motion phenomenology of the perturbed solutions}\label{sec:phenomenology}
The geodesic motion of point particles in spherical symmetric spacetimes is most easily derived from the Lagrangian,
\begin{align}
    2 \mathcal{L} = g_{\mu\nu}\dot x^\mu \dot x^\nu = \mathcal{A}^2 \dot t^2 - \mathcal{B}^2 \dot r^2 - r^2 (\dot \vartheta^2 + \sin^2\vartheta \dot\varphi^2)\,.
\end{align}
There exist two constants of motion, the energy $E= \frac{\partial \mathcal{L}}{\partial \dot t}$ and the angular momentum $L= \frac{\partial \mathcal{L}}{\partial \dot \varphi}$, and, thanks to spherical symmetry, without loss of generality we can restrict the analysis to the equatorial plane $\vartheta = \frac{\pi}{2}$. This results in the fact that the sole remaining equation of motion to solve is
\begin{align}\label{eq:geod}
    \frac{1}{2}\dot r^2 + V(r)  = \frac{1}{2}\dot r^2 + \left(\frac{1}{2 \mathcal{B}^2}\left(\frac{L^2}{r^2}+\sigma\right) - \frac{E^2}{2 \mathcal{A}^2 \mathcal{B}^2}\right) = 0\,,
\end{align}
where the effective potential for the perturbative metric coefficients $\mathcal{A}^2  = (1-\frac{2 M}{r}) + \epsilon a(r)$ and $\mathcal{B}^2  = (1-\frac{2 M}{r})^{-1} + \epsilon b(r)$ is, to first order in $\epsilon$, given by
\begin{align}\label{eq:Veff}
   	V(r) =& - \frac{1}{2} E^2 + \frac{1}{2} \left(1-\frac{2M}{r}\right) \left( \frac{L^2}{r^2} + \sigma \right) \nonumber\\
	&+ \frac{\epsilon}{2} \left[ E^2 \left( \frac{a(r)}{1-\frac{2M}{r}} + b(r)\left(1-\frac{2M}{r}\right)\right) - b(r)\left( \sigma + \frac{L^2}{r^2} \right)\left(1-\frac{2M}{r}\right)^2\right]\,.
\end{align}

We now display some classical observables for our spherically symmetric black whole spacetime, which can be compared to observations in the solar systems, or near black holes, to falsify or find evidence for teleparallel corrections to the geometry of spacetimes. In particular, the combination of the four observables we study allows one to find bounds on the parameters, which label the different theories we study, from observations near black holes.
\vspace{11pt}

Most importantly we will see explicitly that the real and the complex tetrad, as well as the two different real tetrads ($\xi=1$ or $\xi=-1$) lead to very different phenomenology, i.e.\ are constraint as teleparallel corrections to general relativity on a different level.
\vspace{11pt}

This feature of the different tetrads becomes most visible in the perihelion shift, in light deflection/lensing observations and the Shapiro delay. For other observables, like the circular photon orbits, which are the core in the derivation of the shadows of black holes, the different tetrads predict different outcomes but the difference is not as visible as for the aforementioned ones.

To demonstrate these findings we display the explicit expressions for the perihelion shift, the light deflection and the Shapiro delay. Details on their derivation can be found in Appendix~\ref{app:phen}.
\vspace{11pt}

\paragraph{Perihelion Shift:} For the perihelion $\Delta\varphi$ shift of a massive particle ($\sigma=1$) orbit $r=r_c + r_1(\varphi)$ which is a small perturbation $r_1(\varphi)$ of a circular orbit at radius $r_c$, we find:
\begin{itemize}
    \item For the real tetrad \eqref{eq:RealSchw}
    \begin{align}\label{eq:perireal}
    \frac{\Delta \varphi}{\pi} 
    &= \ell  \left(6 - \epsilon\frac{2 (\xi -1)   (4 \alpha +2 \beta +3 \gamma )}{M^2}\right) 
    + \ell ^2 \left(27 - \epsilon \frac{120 (\xi -1)  (\alpha +\beta +\gamma )}{M^2}\right)\nonumber\\
    &+ \ell ^3 \left(135 + \epsilon \frac{ (893 - 887 \xi)\gamma - 896 \alpha (\xi -1) + (890 - 878 \xi)\beta}{M^2}\right)\,.
    \end{align}
    \item For the complex tetrad \eqref{eq:CompSchw} ($q\in \mathbb{N}\geq 1$)
    \begin{align}
        \frac{\Delta \varphi}{\pi}|_{q=1,q>3} &=  \ell  \left(6 + \frac{2 \gamma  \epsilon }{M^2}\right)
        + \ell^2 \left(27 + \epsilon \frac{56 (\beta +\gamma )}{M^2}\right)
        + \ell^3 \left(135 + \epsilon  \frac{(396 \beta +399 \gamma )}{M^2}\right)\, \label{eq:pericmplx1}\,.
    \end{align}
\end{itemize}
For the real tetrad, the sign choice $\xi = 1$ or $\xi = -1$ makes a qualitative difference in the prediction. For the $\xi=1$ tetrad, the changes induced by the teleparallel modifications of GR only become visible at order $\ell^3$, while for the $\xi=-1$ tetrad they already appear at order $\ell$. Similarly, for the complex tetrad, corrections to GR appear already at order $\ell$. Thus, precision measurements of the perihelion shift lead to strong constraints on the teleparallel gravity modification of GR for the $\xi=-1$ and the complex tetrads, while the bounds for the $\xi=1$ tetrad are rather weak.

\vspace{11pt}

\paragraph{Light Deflection:} For the light deflection $\Delta \varphi$, which describes the bending of a null geodesic measured with closest encounter to the central gravitating mass $r_0$, we find up to order $r_0^{-3}$, assuming $r_0>2 M$ :
\begin{itemize}
    \item For the real tetrad \eqref{eq:RealSchw}
    \begin{align}\label{eq:lightdefreal}
        \Delta \varphi = \frac{4 M}{r_0} + \epsilon \left(\frac{(\xi -1)( M (4 (44-9 \pi ) \alpha +8 (29-6 \pi ) \beta +6 (34-7 \pi ) \gamma )+\pi  r_0 (6 \alpha +8 \beta +7 \gamma ) )}{2 \xi  r_0^3}-\frac{\mathfrak{h} 2 M}{r_0}\right)\,.
    \end{align}
    \item For the complex tetrad \eqref{eq:CompSchw}
    \begin{align}
        \Delta\varphi|_{q=2} &= \frac{4 M}{r_0} - \epsilon  \left(\frac{2 \mathfrak{h}  M}{r_0} -\frac{\pi  (3 \alpha +5 \beta +4 \gamma )}{r_0^2} - \frac{2 M ((128-27 \pi ) \alpha +(212-45 \pi ) \beta +2 (85-18 \pi ) \gamma )}{3 r_0^3} \right)\, \label{eq:light_def_cmplx2}\,.
    \end{align}
\end{itemize}
As for the perihelion shift, the predictions for the light deflection angle are very sensitive to the choice of the tetrad. The complex tetrad and the real $\xi=-1$ tetrad lead to way more significant deviations from GR, in comparison to the real $\xi=1$ tetrad. In particular we see that for the former tetrads, in the $M\to0$ limit, i.e. for perturbations of Minkowski spacetime, non-trivial, gravitational effects remain, while for the later tetrad all modifications are proportional to $M$.

\vspace{11pt}
    
\paragraph{Time delays} The traveling time of a light ray emitted at a $r=r_0$ to $r=r_X$ with $r_0>r_X$, from which one can determine the Shapiro Delay,
\begin{align}
    \Delta t_{\textrm{S}}(r_0,r_X) = t(r_0,r_X) - t(r_0,r_X)|_{2 M=0}\,, 
\end{align}
is given by the following, rather lengthy expressions:
\begin{itemize}
    \item For the real tetrad~\eqref{eq:RealSchw}
    \begin{align}\label{eq:shapiroreal}
        t(r_0,r_X) 
        &=  \sqrt{r_X ^2 - r_0 ^2} \left(1 - \frac{\epsilon \mathfrak{h}}{2}\right) - \frac{4\epsilon }{r_0} (\xi -1) (6 \alpha +8 \beta +7 \gamma ) \cos^{-1}\left(\frac{r_0}{r_X}\right)\nonumber \\
        &+ 2M \left[\frac{\sqrt{r_X^2-r_0^2}}{2 (r_0+r_X)} +\ln \left(\frac{ \sqrt{r_X^2-r_0^2}+r_X}{r_0}\right)\right]
        +\frac{M}{2} \mathfrak{h} \epsilon  \left[\frac{3 (r_0-r_X)}{\sqrt{r_X^2-r_0^2}}+2 \ln \left(\frac{r_0}{\sqrt{r_X^2-r_0^2}+r_X}\right)\right]\nonumber\\
        &+ \frac{\epsilon  (\xi -1) 2M}{r_0^2} \Bigg[ \frac{2 (6 \alpha +8 \beta + 7 \gamma ) }{r_X+r_0}\left(\left(r_X+r_0\right)  \left\{\pi-2\tan^{-1}\left(\frac{r_0}{\sqrt{r_X^2-r_0^2}}\right)\right\}  - 2 \sqrt{r_X^2-r_0^2}\right)\nonumber\\
        &-\frac{3 \sqrt{r_X^2-r_0^2}}{r_X} (20 \alpha +26 \beta +23 \gamma ) \Bigg]\,.
    \end{align}
    \item For the complex tetrad \eqref{eq:CompSchw}
    \begin{align}
        t(r_0,r_X)|_{q=2}
        &= \sqrt{r_X ^2 - r_0 ^2} \left(1 - \frac{\epsilon \mathfrak{h}}{2}\right)  + \epsilon \frac{4 (3 \alpha +5 \beta +4 \gamma ) }{r_0}\cos^{-1}\left(\frac{r_0}{r_X}\right) - \epsilon M \mathfrak{h}  \ln \left(\frac{r_0}{\sqrt{r_X^2-r_0^2}+r_X}\right)\nonumber\\
        &+ 2M \Bigg[ \frac{\sqrt{r_X^2-r_0^2}}{2 (r_0+r_X)} +\ln \left(\frac{ \sqrt{r_X^2-r_0^2}+r_X}{r_0}\right) -\epsilon \Bigg\{ \frac{2 (3 \alpha +5 \beta +4 \gamma )}{r_0^2} \left(\pi -2 \tan ^{-1}\left(\frac{r_0}{\sqrt{r_X^2-r_0^2}}\right)\right) \nonumber\\
        & - \frac{\sqrt{r_X^2-r_0^2} }{4 r_0^2 r_X \left(r_X+r_0\right)}\left(4 r_0 (28 \alpha +46 \beta +37 \gamma )+4 r_X (40 \alpha +66 \beta +53 \gamma )-3 \mathfrak{h} r_0^2 r_X\right) \Bigg\} \Bigg]\,.\label{eq:shapirocomp3}
    \end{align}
\end{itemize}
As the perihelion shift and the light deflection angle, the Shapiro time delay has a high sensitivity to the choice of the tetrads. \vspace{11pt}

The findings of this section foster the conclusion that the real $\xi=1$ tetrad is the least constrained by observations for any $f(T,B)$-gravity model, while the $\xi=-1$ and complex tetrad lead to strong constraints on the model parameters.

\section{A brief discussion on time-dependant spherical symmetry}\label{sec:time_depend}
The antisymmetric field equation~\eqref{FE_symm} is explicitly given by
\begin{eqnarray}
    0=\partial_{[\mu}( f_T+f_B)T^{\mu}{}_{\lambda\nu]}=\frac{1}{3}\Big(T_{\lambda} \partial_\nu (f_T+ f_B)-T_{\nu} \partial_\lambda (f_T+f_B)+T^{\mu}{}_{\lambda\nu}\partial_\mu (f_T+f_B)\Big)\,.\label{antifTBphiX}
\end{eqnarray}
If we assume that $T=T(t,r)$ and $\,B=B(t,r)$, all derivatives of the function $f$ would be different to zero. This would mean that for the most general tetrad satisfying spherical symmetry~(\ref{tetrad1}), the only two non-vanishing antisymmetric equation become
\begin{eqnarray}
    0&=&(f_{T,r}+f_{B,r})(C_3 C_5-C_{5,t} C_5-C_6 C_{6,t})+(f_{T,t}+f_{B,t})(C_6 C_{6,r}-C_{4} C_5+C_{5,r} C_5)\,,\\
    0&=&C_6 \Big[C_2 \left(f_{B,t}+f_{T,t}\right)-C_1 \left(f_{B,r}+f_{T,r}\right)\Big]\,,
\end{eqnarray}
where commas represent derivatives. There are two possible ways to solve these equations. The first one is to assume a form of $f$ and replace it in the above equation. Then, one would have two constraint equations. This procedure, of course, depends on the form of $f$ chosen. Another strategy would be to not assume any form of $f$ and assume that all the parenthesis in the above equation are identically zero. Let us concentrate in this way of solving the equations, which gives us the following equations
\begin{eqnarray}
0&=&2 C_3 C_5-2 C_{5,t}C_5-2 C_6 C_{6,t}\,,\\
0&=&2 C_4 C_5-2 C_{5,r}C_5-2 C_6 C_{6,r}\,,\\
0&=&C_2 C_6\,,\quad 0=C_1 C_6\,,\label{confanti}
\end{eqnarray} 
This system is solved only if
\begin{eqnarray}
C_6=0\,,\quad C_3=C_{5,t}\,,\quad C_4=C_{5,r}\,.
\end{eqnarray}
One can easily verify that this solution gives us $T=0$ and then $B=B(t,r)=\lc{R}$. One can then notice that if we impose that the scalars depend on both variables, the antisymmetric field equations are only solved when $f=f(0,B)=f(\lc{R})$, which is a theory where the antisymmetric equations are always satisfied. Thus, one can conclude that the second way of solving the antisymmetric field equations leads to a trivial situation where $T=0$ and the theory just become a one which is fully described by the scalar curvature $\lc{R}$. 

This small example shows that when one has a spacetime depending on two variables, as it is in the time-dependant spherically symmetric case, one cannot solve the equation generically without imposing the form of $f$. This means that in order to solve them, one would need to do the same as one does for the symmetric field equations, which is, to set the form of $f$ and then solve the remaining system of equations for all the tetrad functions appearing in both symmetric and antisymmetric field equations. One can then conclude that it is not possible to show for a generic form of $f$ whether the stationary condition of the Birkhoff's theorem is satisfied in $f(T,B)$ gravity.

\section{Conclusion and outlook\label{sec:conclusion}}

In this work we explore part of the landscape of possible exact and perturbative solutions in teleparallel gravity in the context of spherically symmetric settings. Primarily  the search for these solutions \cite{Paliathanasis:2014iva,Ruggiero:2015oka,DeBenedictis:2016aze,Flathmann:2019khc,Bahamonde:2020vpb,Pfeifer:2021njm,Golovnev:2021htv,Bohmer:2019vff,Boehmer:2020hkn} has been limited to $f(T)$ gravity with the nuances of tetrad-spin connection pair ansatz considerations not being fully exposed. Further, exact solutions have been found but those solutions were only limited to have a dynamics equal to GR plus a cosmological constant (either $f=-T+\Lambda$ or $T=\textrm{const}.$).  In Sec.~\ref{sec:action}, we open this discussion by introducing the need for symmetric and antisymmetric \eqref{FE_symm} field equations which respectively represent the degrees of freedom associated with the tetrad and spin connection. Thus, both sets of equations must be satisfied for a tetrad-spin connection ansatz to produce acceptable solutions.

Another key component to spherically symmetric solutions is that of the role of Bianchi identities since they encode how the equations of motion are related together. General Bianchi identities have been shown for $f(T)$ gravity \cite{Golovnev:2020las}, while in our Sec.~\ref{ssec:BI} we examine the realization of these identities for $f(T,B)$ gravity where the generalized Bianchi identity does lead to the conservation of the energy-momentum tensor when both the symmetric and antisymmetric field equations are satisfied, even if not a usual conserved energy-momentum tensor is used for matter. Of note, the possible addition of a scalar field in this setting would add little to nothing to the identities.

With the concept of tetrad-spin connection pairs satisfying the antisymmetric field equations now established in definitive terms, we then lay out the background of spherical symmetry first in the context of Killing vectors in Sec.~\ref{sec:spherical_symm}. This leads to the general spherically symmetric metric in Eq.~\eqref{metric} but also the tetrad in Eq.~\eqref{sphtetrad} which is the most general spherically symmetric tetrad in the Weitzenb\"{o}ck gauge. Here, the six free $C_i(t,r)$ functions need to be resolved through the equations of motion. A curious feature of this is the property that, provided the metric remains real, these components can take on imaginary components (firstly presented in the literature in~\eqref{tetrad2}) which opens a wide plethora of possible solutions in addition to the intuitive ones. In Sec.~\ref{sec:good_ted_spin} another important property emerges which comes from the antisymmetric set of field equations, namely the branching of solutions due to the vanishing or nonvanishing nature of $C_3$. For the first case where $C_3=0$, we find two solutions of the tetrad dependent on the value of $\xi$ which can have values $\pm1$. This also affects the way that the Minkowski limit is approached (\ref{eq:xi_pos},\ref{eq:xi_neg}) and how the scalars tend to vanish in this limit.

The other branch of the spherically symmetric tetrad, namely $C_3\neq 0$, produces a complex tetrad which at face value may be counter-intuitive. The easiness of the search for exact solutions in the $f(T)$ and $f(T,B)$ gravity case with the complex tetrad is a remarkable property of the model which indeed allowed us to explicitly find some exact solutions, and it is also an interesting mathematical fact in itself.

Note that other complex tetrads have been considered in the literature \cite{Bejarano:2017akj} in the cosmological context, however all these solutions have been found either in vacuum, or for modified gravity coupled to standard baryonic matter. The equations of motion of $f(T,B)$ gravity also admit a physical coupling to the tetrad alone when matter with antisymmetric stress-energy tensor (fermionic matter) is coupled to the antisymmetric part of the equations of motion \cite{Ferraro:2018axk,Bejarano:2019fii}. As our choice of spherically symmetric tetrads exactly vanishes that contribution, it could not be expected that both real and complex tetrad have such coupling. Nonetheless, this issue should be taken with extreme care, since the antisymmetric part of the equations of motion depends purely on the tetrad field, and the coefficients there could vanish and consequently be associated to disappearance of degrees of freedom and therefore related with strong coupling problem \cite{Ferraro:2018tpu,Blagojevic:2020dyq,Golovnev:2020zpv}.

Altogether, there does not seem to be any serious physical problem with using the complex tetrad for representing the spherically symmetric geometry as a background, since the metric (and actually also $T$ and therefore $B$) is purely real. Having said this, we should however make a cautionary remark, especially for those who would like to study perturbations around. For sure, the complex tetrad choice assumes some generalisation of the formulation of the theory. A complex tetrad has 16 complex components, what can be viewed as 32 real variables, with 10 conditions of reality of the metric which still leaves more variables than in the purely real theory. And putting it another way, the complex $f(T)$ equations of motion must be supplemented by equating imaginary parts of the metric components to zero. It is not a priori clear how it would influence the degrees of freedom, nor how well the new equations would interact with the basic ones. It is an open question which should definitely be addressed in future work if we are to take these solutions seriously.

The general setting of an anisotropic perfect fluid is then considered in Sec.~\ref{sec:pert_spherical_sym} where the equations of motion are in Eqs.~(\ref{fieldeqsym1}--\ref{fieldeqsym3}). While exact solutions exist in this expression of the theory \cite{Bahamonde:2019jkf,Golovnev:2021htv}, they are very cumbersome to determine more generally. This becomes all the more difficult when physically motivated solutions are sought. For the case of vacuum solutions in $f(T)$ gravity, an intriguing relation that does not contain the arbitrary Lagrangian is found in Eq.~\eqref{eq:branch1fTB}, leading to the theorem
\begin{theorem}
    In $f(T)$ gravity, only for the case where the model is at most TEGR + Constant, do the $\mathcal{A}(r)$ and $\mathcal{B}(r)$ take on the reciprocal of each other. Moreover, the solution in this case is the Schwarzschild de Sitter solution.
\end{theorem}
It is at this point that we return to the general scenario of $f(T,B)$ gravity where we take perturbations of these ansatz functions about a Schwarzschild background in Eqs.~(\ref{eq:schwazr_per_A})--(\ref{eq:schwazr_per_B}). Before attempting the general solution, we first take a simpler setting of a perturbation about a Minkowski background in Sec.~\ref{sec:min_per_back} which readily provides first order solutions, where we present new perturbed solutions that have not been found before in the literature for the two tetrads. In Sec.~\ref{ssec:pertSchwaReal} we move onto the general Schwarzschild perturbative solution where the tetrad solution readily leads to the metric component solutions in Eqs.~(\ref{eq:Schwar_per_sol_A})--(\ref{eq:Schwar_per_sol_B}). The solution continues to be dependent on $\xi$ for the real tetrad due to the ansatz taken earlier. Together with the conditions of a well behaved Schwarzschild-like horizon \eqref{horizon} and a vanishing of the perturbative components of $\mathcal{B}$ in the asymptotic limit, the two possible value of $\xi$ can be reconsidered. It turns out that while $\xi=1$ produces a well behaved Minkowski limit, the other setting where $\xi=-1$ turns out to be problematic in this area. This seems to be another generic feature of spherically symmetric solutions. It should be noted that we did not prove that the perturbative solutions are indeed convergent to anything physical. This is something to be analysed in the future. And the cases of $\xi=-1$ can raise even more doubts about that since, unlike for $\xi=1$ expansions, there is no guarantee that there really exists a solution anywhere close to what we perturb around.

The second branch of the tetrad ansatz in Eq.~\eqref{tetrad2} is then explored in Sec.~\ref{sec:complex_tetrad} which is the complex solution for the tetrad field in the Weitzenb\"{o}ck gauge. As in the first case, we first take perturbations about a Minkowski background in Sec.~\ref{sec_com_mn_per}, and solve for the first order terms. We then perform a similar calculation for a Schwarzschild background in Sec.~\ref{ssec:pertSchwaIm}, where the well behaved horizon and asymptotic behaviour of $\mathcal{B}$ produce viable solutions. Interestingly, for the instance of $f(T)$ gravity, an exact relationship can be obtained between the metric function \eqref{Bcomplex} in Sec.~\ref{sec:com_f_T_sol}. To this end, we show that exact solutions can be obtained for both $\mathcal{A}$ and the arbitrary $f(T)$ Lagrangian for the $\mathcal{B}={\rm const.}$ scenario. For more realistic spherically symmetric scenarios, we then take a solution similar to Reissner–Nordstr\"{o}m solution in Eq.~\eqref{Acharg} which gives analytic solutions for $\mathcal{B}$ as well as for $f(T)$ (see (\ref{eq:com_f_T_sol1}) and (\ref{eq:com_f_T_sol2})). Analogously, one can also assume a form of $f(T)$ such as the power-law model which results in the metric in Eq.~\eqref{eq:com_f_T_metric}, which has clearly determinable horizons and appropriate asymptotic behaviours. Up to our knowledge, the solutions~\eqref{sol1} and~\eqref{eq:com_f_T_metric} are the first non-trivial exact black hole solutions found in teleparallel gravity beyond TEGR. By non-trivial we refer to solutions that do not have $T=\textrm{const}.$ or $f=T+\Lambda$ which essentially are just solutions in GR plus a cosmological constant. For~\eqref{eq:com_f_T_metric}, it will be interesting to analyse what happens in the interior of the black hole and see if it is possible to have a regular black hole as it was analysed in~\cite{Bohmer:2019vff} by doing numerical and dynamical system techniques. Finally, we close this section with a look into an exact solution for the $f(T,B)=k_1T+F(B)$ model which is a nontrivial $f(B)$ gravity model. By taking a reasonable power-law model for the $\mathcal{B}$ component, we determine an exact form for $\mathcal{A}$ \eqref{eq:com_f_B_solA} which also produces a metric solution with a Schwarzschild-like reciprocal relationship between the $\mathcal{A}$ and $\mathcal{B}$ components \eqref{eq:com_f_B_sol1}.

With the real \eqref{eq:RealSchw} and complex \eqref{eq:CompSchw} tetrad solutions in hand, we can now explore some of the astrophysical phenomenology of these solutions. These observables can provide a crucial step in the realization of the viability of these solutions in terms of the real Universe. To this end, we investigate the perihelion shift which also occurs in
strong field binary systems \eqref{eq:perireal}–\eqref{eq:pericmplx1}, light deflection~\eqref{eq:lightdefreal}-\eqref{eq:light_def_cmplx2} as well as time delay for echo signals~\eqref{eq:shapiroreal}-\eqref{eq:shapirocomp3}. We found that the complex tetrad and the real tetrad with $\xi=-1$ and the complex tetrad lead to corrections to general relativity at lower order than the the real tetrad with $\xi=1$. For completeness we added the phenomena of circular photon orbits~(\ref{eq:circphoreal}), (\ref{eq:circphocmplx1}) and (\ref{eq:circphocmplx2})  which are connected with black hole shadows in the appendix. For this observable no such qualitative difference between the tetrads appears.  These tests will be critical for assess how realistic these novel solutions will be in confronting observational data. Finally, we close this study with a brief, but new, exploration of the fate of time-dependent solutions in TG with Sec.~\ref{sec:time_depend}. There is a long history of attempts at time dependent solutions within TG which have turned out to be very challenging even numerically.

As a future work, it will be interesting to use the complex tetrad as a starting point for constructing axially symmetric solutions as it was done in~\cite{Bahamonde:2020snl}. Since this tetrad gives much simpler field equations in spherical symmetry, a possible generalisation to axial symmetry might give the possibility to find perturbed or exact rotating black hole solutions. Further, one can also use some of the ideas presented in~\cite{Bahamonde:2021qjk} in the context of metric-affine gravity where exact rotating black hole solutions were found (with propagating torsion and nonmetricity).

\begin{acknowledgments}
SB is supported by the Estonian Research Council grants PRG356 “Gauge Gravity” and MOBTT86, and by the European Regional Development Fund CoE program TK133 “The Dark Side of the Universe”. SB also acknowledges JSPS Postdoctoral Fellowships for Research in Japan and KAKENHI Grant-in-Aid for Scientific Research No. JP21F21789. MJG was funded by the Estonian Research Council grant MOBJD622, by the European Regional Development Fund CoE program TK133 “The Dark Side of the Universe” and by FONDECYT-ANID postdoctoral grant 3190531. JLS would like to acknowledge funding support from Cosmology@MALTA which is supported by the University of Malta, as well as the networking support from ``The Malta Council for Science and Technology'' in project IPAS-2020-007. CP was funded by the Deutsche Forschungsgemeinschaft (DFG, German Research Foundation) - Project Number 420243324.
\end{acknowledgments}

\appendix

\section{Perturbations around Minkowski for \texorpdfstring{$h_{(1)}^\mu{}_A$ with $\xi=-1$}{}}\label{sec:othersols}
The form of $b(r)$ can be  expressed in all the solutions as
\begin{equation}
    b(r)=r a'(r)+  2^{3 m-2}\beta \left(m^2-1\right) r^{2-2 m}+2^{3 q-2}\alpha   (q-1) r^{2-2 q}+  2^{3 s+3 w-2}\gamma (s+1) (s+w-1) r^{-2 (s+w-1)}\,.
\end{equation}
Solution for $q=3/2$ and $m\neq3/2,\, s\neq \frac{1}{2}(3-2w)$
\begin{eqnarray}
a(r)&=&a_0-\frac{a_1}{r}+\alpha\frac{4 \sqrt{2}   (\log r+1)}{r}+\beta \frac{  (m (2 m-3)-4) 2^{3 m-4}}{2 m-3} r^{2-2 m}+\gamma\frac{   (s (2 s+2 w-3)-4)2^{3 s+3 w-4}}{2 s+2 w-3} r^{-2 s-2 w+2}\,.\nonumber \\ 
\end{eqnarray}
Solution for $q=m=3/2$ and $s\neq \frac{1}{2}(3-2w)$
\begin{eqnarray}
a(r)&=&a_0-\frac{a_1}{r}+\frac{4 \sqrt{2} (\alpha +\beta )}{r}+\frac{4 \sqrt{2} (\alpha +\beta ) \log r}{r}+\gamma\frac{  (s (2 s+2 w-3)-4)  2^{3 s+3 w-4}}{2 s+2 w-3}r^{-2 s-2 w+2}\,.
\end{eqnarray}
Solution for $q=m=3/2$ and $s= \frac{1}{2}(3-2w)$
\begin{eqnarray}
a(r)&=&a_0-\frac{a_1}{r}+\frac{4 \sqrt{2} (\alpha +\beta +\gamma )}{r}+\frac{4 \sqrt{2}  (\alpha +\beta +\gamma )\log r}{r}\,.
\end{eqnarray}
Solution for $m=3/2$ and $q\neq3/2,\, s\neq \frac{1}{2}(3-2w)$
\begin{eqnarray}
a(r)&=&a_0-\frac{a_1}{r}+\alpha \frac{  2^{3 q-2} }{3-2 q}r^{2-2 q}+\frac{4 \sqrt{2} \beta  (\log r+1)}{r}+\gamma \frac{  \left(2 s^2+s (2 w-3)-4\right)2^{3 s+3 w-4} }{2 s+2 w-3}r^{-2 (s+w-1)}\,.
\end{eqnarray}
Solution for $m=3/2,\, s= \frac{1}{2}(3-2w)$ and $q\neq3/2$
\begin{eqnarray}
a(r)&=&a_0-\frac{a_1}{r}+\alpha\frac{  2^{3 q-2} }{3-2 q}r^{2-2 q}+\frac{4 \sqrt{2} (\beta +\gamma )}{r}+\frac{4 \sqrt{2} (\beta +\gamma ) \log r}{r}\,.
\end{eqnarray}
Solution for $s= \frac{1}{2}(3-2w)$ and $q,m\neq3/2$
\begin{eqnarray}
a(r)&=&a_0-\frac{a_1}{r}+\alpha\frac{  2^{3 q-2} }{3-2q}r^{2-2 q}+\beta \frac{  \left(2 m^2-3 m-4\right) 2^{3 m-4}}{2 m-3}r^{2-2 m}+\frac{4 \sqrt{2} \gamma  (\log r+1)}{r}\,.
\end{eqnarray}

\section{Perturbations around Minkowski for \texorpdfstring{$h_{(2)}^\mu{}_A$}{}}\label{sec:othersols2}
The form of $b(r)$ can be  expressed in all the solutions as
\begin{equation}
    b(r)=r a'(r)+  4^{m-1}\beta \left(2 m^2-m-1\right) r^{2-2 m}+  4^{q-1}\alpha (q-1) r^{2-2 q}+4^{s+w-1}\gamma  (2 s+1)  (s+w-1) r^{-2 s-2 w+2}\,.
\end{equation}
Solution for $q=3/2$ and $m\neq3/2,\, s\neq \frac{1}{2}(3-2w)$
\begin{equation}
a(r)=a_0-\frac{a_1}{r}+\frac{2^{2 m-3}\beta   \left(2 m^2-3 m-2\right) r^{2-2 m}}{2 m-3}+\frac{2^{2 s+2 w-3}\gamma   \left(2 s^2+s (2 w-3)-2\right) r^{-2 s-2 w+2}}{2 s+2 w-3}+\frac{2 \alpha  (\log (r)+1)}{r}\,.
\end{equation}
Solution for $q=m=3/2$ and $s\neq \frac{1}{2}(3-2w)$
\begin{eqnarray}
a(r)&=&a_0-\frac{a_1}{r}+\frac{2^{2 s+2 w-3} \gamma   \left(2 s^2+s (2 w-3)-2\right) r^{-2 s-2 w+2}}{2 s+2 w-3}+\frac{2 (\alpha +\beta ) (\log (r)+1)}{r}\,.
\end{eqnarray}
Solution for $q=m=3/2$ and $s= \frac{1}{2}(3-2w)$
\begin{eqnarray}
a(r)&=&a_0-\frac{a_1}{r}+\frac{2 (\log (r)+1) (\alpha +\beta +\gamma )}{r}\,.
\end{eqnarray}
Solution for $m=3/2$ and $q\neq3/2,\, s\neq \frac{1}{2}(3-2w)$
\begin{eqnarray}
a(r)&=&a_0-\frac{a_1}{r}+\frac{ 4^{q-1}\alpha  r^{2-2 q}}{3-2 q}+\frac{2^{2 s+2 w-3} \gamma  \left(2 s^2+s (2 w-3)-2\right) r^{-2 s-2 w+2}}{2 s+2 w-3}+\frac{2 \beta  (\log (r)+1)}{r}\,.
\end{eqnarray}
Solution for $m=3/2,\, s= \frac{1}{2}(3-2w)$ and $q\neq3/2$
\begin{eqnarray}
a(r)&=&a_0-\frac{a_1}{r}+\frac{4^{q-1}\alpha   r^{2-2 q}}{3-2 q}+\frac{2 (\beta +\gamma ) (\log (r)+1)}{r}\,.
\end{eqnarray}
Solution for $s= \frac{1}{2}(3-2w)$ and $q,m\neq3/2$
\begin{eqnarray}
a(r)&=&a_0-\frac{a_1}{r}+\frac{2^{2 m-3}\beta   \left(2 m^2-3 m-2\right) r^{2-2 m}}{2 m-3}+\frac{ 4^{q-1}\alpha  r^{2-2 q}}{3-2 q}+\frac{2 \gamma  (\log (r)+1)}{r}\,.
\end{eqnarray}

\section{Derivation of Observables}\label{app:phen}
Here we discuss the derivation of the expressions for the circular photon orbits, the perihelion shift, light deflection and Shapiro delay, which we discussed in Section \ref{sec:phenomenology}, following essentially the standard textbook methods, as they can for example be found in \cite{Weinberg:1972kfs}.\vspace{11pt}

\paragraph{Circular Photon Orbits} To find the circular photon orbits we impose $\dot r=0$ and solve $V(r) = 0$ and $V'(r)=0$ with $\sigma=0$ for $r= r_0 + \epsilon r_1$ and $E = E_0 + \epsilon E_1$. For the general perturbation one easily finds
\begin{align}
    r_{\textrm{circ,ph}} = 3M \left( 1 + \epsilon \frac{1}{4} \left( 6M a'(r_0) - 4 a(r_0)\right) \right) \textrm{ and } E = \frac{L}{3 \sqrt{3} M} \left( 1 + \epsilon \frac{3 a(r_0)}{2} \right)\,.
\end{align}
Evaluating this expression yields:
\begin{itemize}
    \item For the real tetrad \eqref{eq:RealSchw}
    \begin{align}\label{eq:circphoreal}
        r_{\textrm{circ,ph}} = 3 M \left( 1 + \epsilon \left(\frac{16 \sqrt{3} \alpha +8 \sqrt{3} \beta +12 \sqrt{3} \gamma -144 (\alpha +\beta +\gamma )}{54 M^2 \xi } +\frac{27 \ln (3) (\alpha +\beta +\gamma )+208 \alpha +232 \beta +220 \gamma }{108 M^2}\right) \right)\,.
    \end{align}
    \item For the complex tetrad \eqref{eq:CompSchw} ($q\in \mathbb{N}\geq 1$)
    \begin{align}
        r_{\textrm{circ,ph}}|_{q=1} &= 3M \left( 1 + \epsilon \frac{ (10 \beta +9 \gamma )}{9 M^2} \right)\,,\label{eq:circphocmplx1}\\
        r_{\textrm{circ,ph}}|_{q>1} &= 3M \left( 1 + \frac{\epsilon}{9 M^2} \left( 10 \beta + 9 \gamma +\frac{\alpha  4^{q-1} 9^{2-q} q M^{4-2 q}}{2 q-3}\right)\right)\label{eq:circphocmplx2}\,.
    \end{align}
\end{itemize}

\vspace{12pt}

\paragraph{Perihelion Shift:} The perihelion $\Delta\varphi$ shift of a massive particle ($\sigma=1$) orbit $r=r_c + r_1(\varphi)$ which is a small perturbation $r_1(\varphi)$ of a circular orbit at radius $r_c$, is given by, see for example \cite{Bahamonde:2019zea},
\begin{align}
    \Delta \varphi =2\pi \left(\frac{L}{r_{\text{c}}^2\sqrt{V''(r_{\text{c}})}}-1\right)\,,
\end{align}
where $L$ is the angular momentum of the circular orbit $r_c$. Using the perturbative potential (\ref{eq:Veff}), splitting $L=L_0+\epsilon L_1$ and $E=E_0+\epsilon E_1$ with $L_0=\frac{r_c \sqrt{2 M}}{\sqrt{2 r_c-3 2 M}}$, $L_1 = \frac{r_c^2 \left(r_c (r-2 M) a'(r_c)- 2 M a(r_c)\right)}{\sqrt{2 M} (2 r_c-3 2 M)^{3/2}}$ and $E_0 = \frac{r_c-2 M}{\sqrt{r_c^2-\frac{3 r_c 2 M}{2}}}$, and introducing the dimensionless parameter $\ell = \frac{M}{r_c}$ we find to first order in $\epsilon$
\begin{align}
    \Delta \varphi 
    &= 2 \pi  \left(\frac{1}{\sqrt{1-6 \ell}}-1\right) \\
    &+ \frac{\pi  \epsilon  \left((2 \ell-1) \left(2 M \left(4 \ell (1-4 \ell) a'(\tfrac{M}{\ell})+(1-2 \ell) 2 M a''(\tfrac{M}{\ell})\right)-8 \ell^3 (4 \ell (3 \ell-2)+1) b(\tfrac{M}{\ell})\right)-32 \ell^4 a(\tfrac{M}{\ell})\right)}{8 \sqrt{1-6 \ell} \ell^3 (4 \ell (3 \ell-2)+1)}\,.
\end{align}
For weak gravity, for example in solar system tests, one assumes $\ell$ to be small and performs a power series expansion in $\ell$. This cannot be done in general since the dependence on $\ell$ depends highly on the perturbation functions $a$ and~$b$. The results, up to order $\ell^3$, are:
\begin{itemize}
    \item For the real tetrad \eqref{eq:RealSchw}
    \begin{align}
    \frac{\Delta \varphi}{\pi} 
    &= \ell  \left(6 - \epsilon\frac{2 (\xi -1)   (4 \alpha +2 \beta +3 \gamma )}{M^2}\right) 
    + \ell ^2 \left(27 - \epsilon \frac{120 (\xi -1)  (\alpha +\beta +\gamma )}{M^2}\right)\nonumber\\
    &+ \ell ^3 \left(135 + \epsilon \frac{ (893 - 887 \xi)\gamma - 896 \alpha (\xi -1) + (890 - 878 \xi)\beta}{M^2}\right)\,.
    \end{align}
    \item For the complex tetrad \eqref{eq:CompSchw} ($q\in \mathbb{N}\geq 1$)
    \begin{align}
        \frac{\Delta \varphi}{\pi}|_{q=1,q>3} &=  \ell  \left(6 + \frac{2 \gamma  \epsilon }{M^2}\right)
        + \ell^2 \left(27 + \epsilon \frac{56 (\beta +\gamma )}{M^2}\right)
        + \ell^3 \left(135 + \epsilon  \frac{(396 \beta +399 \gamma )}{M^2}\right)\,,\\
        \frac{\Delta \varphi}{\pi}|_{q=2} &= \ell  \left(6 + \epsilon \frac{2(2 \alpha +  \gamma)}{M^2}\right) 
        + \ell^2 \left(27 + \epsilon  \frac{56 (\alpha + \beta +\gamma)}{M^2}\right)
        + \ell^3 \left(135 + \epsilon  \frac{(402 \alpha + 396 \beta +399 \gamma) }{M^2}\right)\,,\\
        \frac{\Delta \varphi}{\pi}|_{q=3} &= \ell  \left(6 + \epsilon \frac{2\gamma}{M^2}\right) 
        + \ell^2 \left(27 + \epsilon  \frac{56 (\beta +\gamma)}{M^2}\right)
        + \ell^3 \left(135 + \epsilon  \frac{32 \alpha + M^2(396 \beta +399 \gamma) }{M^4}\right)\,.
    \end{align}
\end{itemize}

\vspace{11pt}

\paragraph{Light Deflection:} Light deflection is described by the bending of a null geodesic measured in standard spherical coordinates
\begin{align}
    \Delta \varphi = 2 \int_{r_0}^\infty \left(\frac{L}{r^2 \sqrt{-2V(r)}} d r \right) - \pi
\end{align}
where $r_0$ is the closest encounter of the light ray to the central mass object. At this point $\dot r = 0$ and we can use $r_0$ to express the energy and angular momentum as $L^2=(E^2 \mathcal{A}^{-2}-\sigma)r_0$. Considering the perturbation function $a$ and $b$ as function of $r$ and $M$, by setting $a(r)=a(r,r_s)$ and $b(r)=b(r,r_s)$, where $r_s=2M$, we can display the integrand of the light deflection angle to first order in $\epsilon$ and $2 M$, as
\begin{align}
    \frac{L}{r^2 \sqrt{-2V(r)}}
    &= \frac{r_0}{r \sqrt{r^2-r_0^2}} + \frac{2 M \left(r^2+r r_0+r_0^2\right)}{2 r^2 (r+r_0) \sqrt{r^2-r_0^2}}\\
    &+\epsilon  \left(\frac{2 M \left(-2 r^2 a(r,0)+2 r r_0 (r+r_0) \partial_{r_s}b(r,0)+\left(r^2-r r_0-r_0^2\right) b(r,0)\right)}{4 r^2 (r+r_0) \sqrt{r^2-r_0^2}}+\frac{r_0 b(r,0)}{2 r \sqrt{r^2-r_0^2}}\right)\,.
\end{align}
The integration of these terms can then be done for the different models. Assuming $r_0>2 M$ we find up to order $r_0^{-3}$:
\begin{itemize}
    \item For the real tetrad \eqref{eq:RealSchw}
    \begin{align}
        \Delta \varphi = \frac{4 M}{r_0} + \epsilon \left(\frac{(\xi -1)( M (4 (44-9 \pi ) \alpha +8 (29-6 \pi ) \beta +6 (34-7 \pi ) \gamma )+\pi  r_0 (6 \alpha +8 \beta +7 \gamma ) )}{2 \xi  r_0^3}-\frac{\mathfrak{h} 2 M}{r_0}\right)\,.
    \end{align}
    \item For the complex tetrad \eqref{eq:CompSchw}
    \begin{align}
        \Delta\varphi|_{q=1} &= \frac{4 M}{r_0} - \epsilon \left( \frac{2 \mathfrak{h}  M}{r_0} - \frac{\pi  (5 \beta +4 \gamma )}{r_0^2} - \frac{2 M ((212-45 \pi ) \beta - 2 (85-18 \pi ) \gamma )}{3 r_0^3} \right)\,,\\
        \Delta\varphi|_{q=2} &= \frac{4 M}{r_0} - \epsilon  \left(\frac{2 \mathfrak{h}  M}{r_0} -\frac{\pi  (3 \alpha +5 \beta +4 \gamma )}{r_0^2} - \frac{2 M ((128-27 \pi ) \alpha +(212-45 \pi ) \beta +2 (85-18 \pi ) \gamma )}{3 r_0^3} \right)\,,\\
        \Delta\varphi|_{q\geq3} &= \frac{4 M}{r_0} - \epsilon  \left(\frac{2 \mathfrak{h}  M}{r_0} + \frac{2 M (45 \pi  \beta -212 \beta +36 \pi  \gamma -170 \gamma )}{3 r_0^3}\right)\,.
    \end{align}
\end{itemize}

\vspace{11pt}
    
\paragraph{Time delays} The traveling time of a light ray emitted at a $r=r_0$ to $r=r_X$ with $r_0>r_X$ is determined by the integral
\begin{align}
    t(r_X,r_0) 
    &= \int_{r_0}^{r_X} \left(\frac{E}{\mathcal{A}^2\sqrt{-2 V(r)}}\right) d r\,,
\end{align}
where, again to first order in $\epsilon$, $2 M$ and $\epsilon 2 M$, the integrand is given by
\begin{align}
    \frac{E}{\mathcal{A}\sqrt{-2 V(r)}}
    &= \frac{r}{\sqrt{r^2-r_0^2}} +  \frac{( 2 r + 3 r_0 ) M}{(r+r_0) \sqrt{r^2-r_0^2}}\\
    &+\epsilon  \left(\frac{r ( b(r,0) - a(r,0) )}{2 \sqrt{r^2-r_0^2}} - 2 M \frac{2 r ( r + r_0 ) \left(\partial_{r_s}a(r,0)-\partial_{r_s}b(r,0)\right) + (4 r + 7 r_0) a(r,0)- r_0 b(r,0)}{4 (r+r_0) \sqrt{r^2-r_0^2}} \right)\,.
\end{align}
The difference between the time of flight of the light ray for $2 M\neq0$ and $2 M=0$ is called the Shapiro Delay
\begin{align}
    \Delta t_{\textrm{S}}(r_0,r_X) = t(r_0,r_X) - t(r_0,r_X)|_{2 M=0}\,. 
\end{align}
The integration of these terms can again be done for the different models. Assuming $r_0>2 M$ we find:
\begin{itemize}
    \item For the real tetrad~\eqref{eq:RealSchw}
    \begin{align}
        t(r_0,r_X) 
        &=  \sqrt{r_X ^2 - r_0 ^2} \left(1 - \frac{\epsilon \mathfrak{h}}{2}\right) - \frac{4\epsilon }{r_0} (\xi -1) (6 \alpha +8 \beta +7 \gamma ) \cos^{-1}\left(\frac{r_0}{r_X}\right)\nonumber \\
        &+ 2M \left[\frac{\sqrt{r_X^2-r_0^2}}{2 (r_0+r_X)} +\ln \left(\frac{ \sqrt{r_X^2-r_0^2}+r_X}{r_0}\right)\right]
        +\frac{M}{2} \mathfrak{h} \epsilon  \left[\frac{3 (r_0-r_X)}{\sqrt{r_X^2-r_0^2}}+2 \ln \left(\frac{r_0}{\sqrt{r_X^2-r_0^2}+r_X}\right)\right]\nonumber\\
        &+ \frac{\epsilon  (\xi -1) 2M}{r_0^2} \Bigg[ \frac{2 (6 \alpha +8 \beta + 7 \gamma ) }{r_X+r_0}\left(\left(r_X+r_0\right)  \left\{\pi-2\tan^{-1}\left(\frac{r_0}{\sqrt{r_X^2-r_0^2}}\right)\right\}  - 2 \sqrt{r_X^2-r_0^2}\right)\nonumber\\
        &-\frac{3 \sqrt{r_X^2-r_0^2}}{r_X} (20 \alpha +26 \beta +23 \gamma ) \Bigg]\,.
    \end{align}
    \item For the complex tetrad \eqref{eq:CompSchw}
    \begin{align}
        t(r_0,r_X)|_{q=1}
        &=  \sqrt{r_X ^2 - r_0 ^2} \left(1 - \frac{\epsilon \mathfrak{h}}{2}\right) +  \epsilon \frac{4  (5 \beta +4 \gamma ) }{r_0}\cos^{-1}\left(\frac{r_0}{r_X}\right)\nonumber \\
        &+ 2M \Bigg[ \frac{\sqrt{r_X^2-r_0^2}}{2 (r_0+r_X)} +\ln \left(\frac{ \sqrt{r_X^2-r_0^2}+r_X}{r_0}\right)
        - \epsilon \Bigg\{\frac{2 (5 \beta +4 \gamma )}{r_0^2} \left(\pi -2 \tan ^{-1}\left(\frac{r_0}{\sqrt{r_X^2-r_0^2}}\right)\right) \nonumber \\
        &-\frac{\sqrt{r_X^2-r_0^2} }{4 r_0^2 r_X \left(r_X+r_0\right)}\left(4 r_0 (46 \beta +37 \gamma )+r_X \left(264 \beta +212 \gamma -3 \mathfrak{h} r_0^2\right)\right)
        - \frac{1}{2} \mathfrak{h}  \ln \left(\frac{r_0}{\sqrt{r_X^2-r_0^2}+r_X}\right) \Bigg\}\Bigg]\,,\\
        t(r_0,r_X)|_{q=2}
        &= \sqrt{r_X ^2 - r_0 ^2} \left(1 - \frac{\epsilon \mathfrak{h}}{2}\right)  + \epsilon \frac{4 (3 \alpha +5 \beta +4 \gamma ) }{r_0}\cos^{-1}\left(\frac{r_0}{r_X}\right) - \epsilon M \mathfrak{h}  \ln \left(\frac{r_0}{\sqrt{r_X^2-r_0^2}+r_X}\right)\nonumber\\
        &+ 2M \Bigg[ \frac{\sqrt{r_X^2-r_0^2}}{2 (r_0+r_X)} +\ln \left(\frac{ \sqrt{r_X^2-r_0^2}+r_X}{r_0}\right) -\epsilon \Bigg\{ \frac{2 (3 \alpha +5 \beta +4 \gamma )}{r_0^2} \left(\pi -2 \tan ^{-1}\left(\frac{r_0}{\sqrt{r_X^2-r_0^2}}\right)\right) \nonumber\\
        & - \frac{\sqrt{r_X^2-r_0^2} }{4 r_0^2 r_X \left(r_X+r_0\right)}\left(4 r_0 (28 \alpha +46 \beta +37 \gamma )+4 r_X (40 \alpha +66 \beta +53 \gamma )-3 \mathfrak{h} r_0^2 r_X\right) \Bigg\} \Bigg]\,.
    \end{align}
\end{itemize}

\bibliographystyle{utphys}
\bibliography{references}

\end{document}